\documentclass[12pt]{article}

\usepackage[margin=1in]{geometry}
\usepackage{setspace}
\usepackage[T1]{fontenc} % correct text layer for accented characters
\usepackage{lmodern}
\usepackage[english]{babel}
\usepackage{amsmath, amssymb}
\usepackage{amsthm}
\usepackage{mathtools}
\usepackage{bm}
\usepackage{graphicx}
\usepackage{enumitem}
\usepackage{needspace}
\usepackage{subcaption}
\usepackage{tikz}
\usepackage{pgfplots}
\pgfplotsset{compat=1.18}
\usetikzlibrary{arrows.meta}

\usepackage[square,semicolon]{natbib}
\usepackage{hyperref}
\hypersetup{colorlinks=true,citecolor=blue,linkcolor=blue,urlcolor=blue,breaklinks=true,
    pdftitle={Why Is the European Electricity Auction Often in Equilibrium? Duality Gaps in Partially Nonconvex Optimization},
    pdfauthor={Thomas H\"ubner}}
\usepackage{cleveref}
\usepackage{xurl} % allows line breaks anywhere in URLs (reference list)

\newtheorem{theorem}{Theorem}
\newtheorem{lemma}{Lemma}
\newtheorem{proposition}{Proposition}
\newtheorem{corollary}{Corollary}

\theoremstyle{definition}
\newtheorem{definition}{Definition}
\newtheorem{assumption}{Assumption}
\newtheorem{example}{Example}
\newtheorem{remark}{Remark}

\crefname{assumption}{Assumption}{Assumptions}
\Crefname{assumption}{Assumption}{Assumptions}

\renewcommand{\emph}[1]{\textit{#1}}

\providecommand{\keywords}[1]{\small\textbf{\textit{Keywords---}} #1}

\DeclareMathOperator*{\argmax}{arg\,max}
\DeclareMathOperator*{\argmin}{arg\,min}

\newcommand{\relinterior}[1]{\operatorname{relint}(#1)}
\newcommand{\cl}[1]{\operatorname{cl}(#1)}

\newcommand{\dom}[1]{\operatorname{dom}(#1)}
\newcommand{\im}[1]{\operatorname{im}(#1)}
\newcommand{\epi}[1]{\operatorname{epi}(#1)}
\newcommand{\hyp}[1]{\operatorname{hyp}(#1)}
\newcommand{\cav}[1]{\operatorname{cav}(#1)}
\newcommand{\co}[1]{\operatorname{co}(#1)}
\newcommand{\ext}[1]{\operatorname{ext}(#1)}
\newcommand{\var}[1]{\operatorname{Var}(#1)}
\newcommand{\E}[1]{\mathbb{E}\left[#1\right]}
\newcommand{\card}[1]{\left| #1 \right|}
\DeclarePairedDelimiter{\norm}{\lVert}{\rVert}

\definecolor{cbMidnightBlue}{HTML}{08254b} % For function curves and main text
\definecolor{cbDarkNavy}{HTML}{084594}     % For true epigraph shading
\definecolor{cbDeepSteel}{HTML}{2171b5}    % For convex hull extension
\definecolor{cbDomainBlue}{HTML}{023858}   % For domain highlights

\title{Why Is the European Electricity Auction\\ Often in Equilibrium?\\[0.0em] {Duality Gaps in Partially Nonconvex Optimization}}

\author{Thomas H\"ubner\thanks{Thomas H\"ubner (thuebner@ethz.ch) is with the Department of Information Technology and Electrical Engineering, ETH Z\"urich. This work was supported by the Swiss Federal Office of Energy's SWEET program under the PATHFNDR project (Grant No. SI/502259). This article significantly extends the first version of this preprint, titled ``Approximate Equilibria in Nonconvex Markets: Theory and Evidence from European Electricity Auctions'' (2025).}}

\date{November 2026}

\begin{document}
%%%%%%%%%%%%%%%%

\maketitle

\begin{abstract}
The European electricity auction determines prices and allocations daily by solving a mixed-integer welfare maximization problem. Analyzing data from nine countries in 2023, we find that its duality gap was frequently zero (on over 80\% of days in three of them), which means that a Walrasian equilibrium existed despite the nonconvexity. To explain this observation, we develop a theory of duality gaps in partially nonconvex optimization, that is, separable problems with both convex and nonconvex functions. Assuming that the convex hulls of the functions' epigraphs are independent and identically distributed, we derive lower bounds on the distribution of the duality gap. They show that zero or small gaps become increasingly likely as convex functions outnumber nonconvex ones, which is the case in the auction. For the special case of welfare maximization, proving these bounds via Walrasian equilibria reveals the economic reasons behind them.
\end{abstract}

\noindent\keywords{duality gap, Shapley--Folkman lemma, partially nonconvex optimization, mixed-integer programming, Walrasian equilibrium, electricity auctions}

\smallskip
\noindent{\small\textbf{\textit{MSC2020---}} Primary: 90C46; Secondary: 90C11, 91B26, 91B50}

\vspace{1cm}

\section{Introduction}

Every day, the European day-ahead electricity auction determines prices, flows, production, and consumption of electricity across most European countries by solving a welfare maximization problem~\citep{madani2015computationally}. This problem is formulated as a mixed-integer program: it combines thousands of convex bids, which give rise to continuous variables, with far fewer nonconvex bids, which give rise to integer variables. Economically, a zero duality gap of this problem implies that a Walrasian equilibrium exists in the market, which is typically not expected in nonconvex markets~\citep{baldwin2019understanding,tran2019product}. Yet, in data spanning 281 days in 2023 across nine countries, the duality gap was zero on more than 80\% of days in Austria, Switzerland, and Poland, whereas in France, Germany, and Great Britain, it was zero on fewer than 20\% of days. As \Cref{fig: zero gap vs convexity} shows, the duality gap tends to be zero more often the larger the share of convex bids in the total bid volume.

% Data: results.csv (281 days, 9 countries).
%   y: share of days with zero duality gap, i.e., |absolute lost opportunity cost| <= 1e-6
%      (reproduces tab:equilibria exactly).
%   x: median over days of convex volume / (convex volume + nonconvex volume), in %
%      (row "Share of total bid volume" in subtab: simple bid curves).
%   Spearman rank correlation: 0.83.
\begin{figure}[t]
\centering
\begin{tikzpicture}
    \begin{axis}[
        width=0.78\linewidth, height=0.44\linewidth,
        scale only axis,
        xmin=60, xmax=100,
        xtick={60,70,80,90,100},
        ymin=0, ymax=100,
        ytick={0,20,40,60,80,100},
        axis lines=left,
        axis line style={-, thin},
        tick style={thin, black},
        major tick length=3pt,
        ymajorgrids, xmajorgrids,
        grid style={black!12, thin},
        tick label style={font=\small},
        label style={font=\small},
        xlabel={Convex share of bid volume (\%)},
        ylabel={Days with zero duality gap (\%)},
        clip=false,
    ]
        \addplot[only marks, mark=*, mark size=2.4pt, color=cbDarkNavy] coordinates {
            (63.612, 5.338)  % GB
            (66.221, 38.434) % BE
            (68.185, 21.352) % NL
            (80.574, 13.167) % DE
            (81.602, 16.370) % FR
            (89.751, 59.075) % FI
            (93.286, 85.053) % AT
            (96.755, 82.206) % CH
            (99.852, 99.288) % PL
        };
        \begin{scope}[font=\footnotesize, inner sep=3pt]
            \node[anchor=west]       at (axis cs:63.612, 5.338)  {GB};
            \node[anchor=west]       at (axis cs:66.221, 38.434) {BE};
            \node[anchor=west]       at (axis cs:68.185, 21.352) {NL};
            \node[anchor=north east] at (axis cs:80.574, 13.167) {DE};
            \node[anchor=south west] at (axis cs:81.602, 16.370) {FR};
            \node[anchor=west]       at (axis cs:89.751, 59.075) {FI};
            \node[anchor=south east] at (axis cs:93.286, 85.053) {AT};
            \node[anchor=west]       at (axis cs:96.755, 82.206) {CH};
            \node[anchor=east]       at (axis cs:99.852, 99.288) {PL};
        \end{scope}
    \end{axis}
\end{tikzpicture}
\caption{Frequency of a zero duality gap versus the convex share of the bid volume in nine countries. The vertical axis shows the share of the 281 days on which the duality gap was zero, i.e., a Walrasian equilibrium existed (cf.~\Cref{tab:equilibria}); each country is analyzed with its net position fixed (\Cref{subsec: duality gaps}). The horizontal axis shows the median daily share of convex simple bids in the total bid volume (cf.~\Cref{tab: bid formats}). The Spearman rank correlation between the two quantities is 0.83.}\label{fig: zero gap vs convexity}
\end{figure}

This pattern is intuitive: the smaller the nonconvex part of a problem relative to its convex part, the smaller one would expect its duality gap to be. Classical bounds on the duality gap, however, cannot account for it, as they depend solely on the nonconvex part of a problem~\citep{aubin1976estimates,cook1986sensitivity}: in the worst case, adding convex bids does not reduce the duality gap at all. Explaining why the auction is so often in equilibrium therefore requires a theory that goes beyond the worst case and captures the influence of the convex part.

In this article, we develop such a theory by studying duality gaps in partially nonconvex optimization, i.e., in separable optimization problems that contain both convex and nonconvex functions. In a probabilistic setting, we derive lower bounds on the distribution of the duality gap, which show that zero or small duality gaps become increasingly likely as convex functions outnumber nonconvex ones. For the important special case of welfare maximization problems, we present alternative proofs that exploit the connection between duality gaps and the existence of Walrasian equilibria and thereby reveal the economic reasons behind this effect. Finally, we return to the European electricity auction and discuss its empirical duality gaps in light of this newly developed theory.

The rest of this article is structured as follows. In the remainder of this introduction, we introduce partially nonconvex optimization problems and summarize our theoretical results in \Cref{subsec: partially nonconvex}, before giving an illustrative example followed by a literature review in \Cref{sec: example,subsec: literature review}. The main part of the article then begins with \Cref{sec: duality gap}, which derives lower bounds on the probability distribution of the duality gap of partially nonconvex problems. \Cref{sec: equilibrium} focuses on the special case of welfare maximization problems, offering an alternative proof with greater economic insight. \Cref{sec: electricity} returns to the European electricity auction, presents the empirical duality gap distribution of its welfare maximization problem, and discusses it in light of the results from \Cref{sec: duality gap,sec: equilibrium}. Finally, \Cref{sec: conclusions} concludes the article.

\subsection{Partially Nonconvex Optimization}\label{subsec: partially nonconvex}

Formally, we study the duality gap $v^{\ast} - v^{\ast\ast}$ of the separable optimization problem
\begin{subequations}\label{eq: problem}
    \begin{align}
        v^\ast := \;\; \inf_{(x_i)_{i\in\mathcal{I}}} & \quad \sum_{i\in\mathcal{I}} f_i^0(x_i)  \\
        \text{s.t.} & \quad \sum_{i\in\mathcal{I}} f_i^k(x_i) \le 0 \quad \forall k\in\mathcal{K}
    \end{align}
\end{subequations}
and its Lagrange dual
\begin{equation}\label{eq: dual}
    v^{\ast\ast} := \;\; \sup_{\lambda\ge 0} \; \sum_{i\in\mathcal{I}} \; \inf_{x_i} \; \left[ f_i^0(x_i)  + \sum_{k\in\mathcal{K}} \lambda_k f_i^k(x_i) \right] ,
\end{equation}
where $f_i^k:\mathbb{R}^{n}\to\mathbb{R}\cup\{\infty\}$ with $i\in\mathcal{I}=\{1,\ldots,I\}$ and $\mathcal{K}=\{1,\ldots,m\}$. 
 
We write $\mathcal{K}_0:=\{0\}\cup\mathcal{K}$, and let $f_i=(f_i^k)_{k\in\mathcal{K}_0}$ be the vector-valued function associated with index $i$.
Its epigraph is given by $\epi{f_i} :=\big\{ (x_i,z_i)\in\mathbb{R}^{n+m+1} \mid x_i\in\mathbb{R}^n, \; z_i^k \ge f_i^k(x_i) \; \forall k\in\mathcal{K}_0 \big\}$.

The dual~\eqref{eq: dual} relaxes only the coupling constraints of~\eqref{eq: problem}; restrictions encoded in the domains of the functions $f_i$, such as integrality, remain in the inner minimizations.
 
We assume the following:
\begin{itemize}
    \item \textit{Partially Nonconvex}: For indices $i\in\mathcal{C}\subseteq\mathcal{I}$, the epigraphs $\epi{f_i}$ are convex. For indices $i\in\mathcal{N}=\mathcal{I}\setminus\mathcal{C}$, the epigraphs $\epi{f_i}$ might be nonconvex.
    \item \textit{Random}: For all indices $i\in\mathcal{I}$, the convex hulls of the epigraphs $\co{\epi{f_i}}$ are independent and identically distributed (i.i.d.).
\end{itemize}

The reason we introduce randomness into the optimization model is that, without it, we cannot understand the influence of the number of convex functions $\card{\mathcal{C}}$ on the duality gap $v^{\ast} - v^{\ast\ast}$: as shown by \citet{aubin1976estimates}, they do not matter in the worst case. For instance, one can add an arbitrary number of convex functions $f_i \equiv 0$ that map all values to zero, and thus do not influence the duality gap at all. By introducing randomness into the problem, we can go beyond such worst-case instances and instead ask what duality gap we can \textit{expect on average}.
The assumption that the convex hulls are i.i.d. introduces this necessary randomness into the optimization problem while still allowing us to distinguish between convex and nonconvex functions. (Note that assuming all functions $f_i$ to be i.i.d. would prevent us from distinguishing between convex and nonconvex functions.)
 
\begin{figure}[t]
\centering
\begin{tikzpicture}[scale=1]
    \begin{axis}[
        width=10cm, height=7cm,
        xmin=-0.5, xmax=5.5,
        ymin=-0.65, ymax=4.5,
        axis lines=middle,
        inner axis line style={-Stealth},
        xlabel={$x_i$}, ylabel={},
        ticks=none,
        axis line style={thick}
    ]
        % -------------------------------------------------------------
        % 1. DEFINE KEY COORDINATES FOR THE GEOMETRY
        % -------------------------------------------------------------
        % Segment 1: Domain [0.5, 2.0]
        \coordinate (Start1) at (axis cs: 0.5, 3.5);
        \coordinate (Min1)   at (axis cs: 1.2, 1.2); % Minimum of piece 1
        \coordinate (End1)   at (axis cs: 2.0, 2.5);
        % Segment 2: Domain [3.2, 4.8]
        \coordinate (Start2) at (axis cs: 3.2, 2.2);
        \coordinate (Min2)   at (axis cs: 4.0, 1.0); % Minimum of piece 2
        \coordinate (End2)   at (axis cs: 4.8, 3.2);
        % -------------------------------------------------------------
        % 2. SHADING REGIONS (Layers: Background to Foreground)
        % -------------------------------------------------------------
        % Base Layer: Fill the entire Convex Hull area up to the top ceiling
        % It uses the flat bridge connecting Min1 and Min2 at the bottom.
        \fill[cbDeepSteel!25] (Start1) .. controls (axis cs:0.8,2.2) and (axis cs:1.0,1.5) .. (Min1)
                         -- (Min2)
                         .. controls (axis cs:4.3,1.7) and (axis cs:4.6,2.5) .. (End2)
                         -- (axis cs:4.8,4.2) -- (axis cs:0.5,4.2) -- cycle;
        % Top Layer: Fill the True Epigraph (epi f_i) over valid domain pieces
        % Segment 1 epigraph
        \fill[cbDarkNavy!45] (Start1) .. controls (axis cs:0.8,2.2) and (axis cs:1.0,1.5) .. (Min1)
                               .. controls (axis cs:1.4,1.0) and (axis cs:1.7,1.8) .. (End1) 
                      -- (axis cs:2.0,4.2) -- (axis cs:0.5,4.2) -- cycle;
        % Segment 2 epigraph
        \fill[cbDarkNavy!45] (Start2) .. controls (axis cs:3.4,1.8) and (axis cs:3.7,1.2) .. (Min2)
                               .. controls (axis cs:4.3,1.7) and (axis cs:4.6,2.5) .. (End2) 
                      -- (axis cs:4.8,4.2) -- (axis cs:3.2,4.2) -- cycle;
        % -------------------------------------------------------------
        % 3. DRAW LINES & CURVES
        % -------------------------------------------------------------
        % Draw the true nonconvex function segments
        \draw[ultra thick, cbMidnightBlue] (Start1) .. controls (axis cs:0.8,2.2) and (axis cs:1.0,1.5) .. (Min1)
                                                  .. controls (axis cs:1.4,1.0) and (axis cs:1.7,1.8) .. (End1);
        \draw[ultra thick, cbMidnightBlue] (Start2) .. controls (axis cs:3.4,1.8) and (axis cs:3.7,1.2) .. (Min2)
                                                  .. controls (axis cs:4.3,1.7) and (axis cs:4.6,2.5) .. (End2);
        \node[black, right] at (axis cs:4.4, 1.8) {$f_i(x_i)$};
        % CORRECTED: The Convex Hull boundary line connecting the two minima
        \draw[thick, dashed, cbDomainBlue] (Min1) -- (Min2);
        % -------------------------------------------------------------
        % 4. DOMAIN REPRESENTATION ON X-AXIS
        % -------------------------------------------------------------
        \draw[line width=2.0pt, cbDomainBlue] (axis cs:0.5,0) -- (axis cs:2.0,0);
        \draw[line width=2.0pt, cbDomainBlue] (axis cs:3.2,0) -- (axis cs:4.8,0);
        % Interval brackets
        \node[cbDomainBlue, font=\bfseries] at (axis cs:0.5,0) {$[$};
        \node[cbDomainBlue, font=\bfseries] at (axis cs:2.0,0) {$]$};
        \node[cbDomainBlue, font=\bfseries] at (axis cs:3.2,0) {$[$};
        \node[cbDomainBlue, font=\bfseries] at (axis cs:4.8,0) {$]$};
        % Domain Labels
        \node[below, black] at (axis cs:1.25,-0.1) {$\dom{f_i}$};
        \node[below, black] at (axis cs:4.0,-0.1) {$\dom{f_i}$};
        % Vertical guidelines from domain boundaries
        \draw[dotted, gray!70] (Start1) -- (axis cs:0.5,0);
        \draw[dotted, gray!70] (End1) -- (axis cs:2.0,0);
        \draw[dotted, gray!70] (Start2) -- (axis cs:3.2,0);
        \draw[dotted, gray!70] (End2) -- (axis cs:4.8,0);
        % -------------------------------------------------------------
        % 5. TEXT ANNOTATIONS FOR THE SHADED REGIONS
        % -------------------------------------------------------------
        \node[black, right] at (axis cs:0.8, 3.0) {$\epi{f_i}$};
        \node[black, right] at (axis cs:3.5, 2.8) {$\epi{f_i}$};
        \node[black, right] at (axis cs:1.8, 1.6) {$\co{\epi{f_i}}$};
    \end{axis}
\end{tikzpicture}
\caption{The epigraph $\epi{f_i}$ and its convex hull $\co{\epi{f_i}}$.}\label{fig:nonconvex_domain_epigraph}
\end{figure}

In this random setting, we obtain two results on the distribution of the duality gap $v^{\ast} - v^{\ast\ast}$. \Cref{theorem: zero duality gap} gives a lower bound on the probability of a zero duality gap, depending on the number of convex functions $\card{\mathcal{C}}$, the number of constraints $m$, and the total number of functions $\card{\mathcal{I}}$. As $\card{\mathcal{C}}$ increases, this bound converges to 1, showing that a zero duality gap becomes increasingly likely as the problem becomes more convex, in line with the pattern in \Cref{fig: zero gap vs convexity}.

\Cref{theorem: non-zero duality gap} extends this to a lower bound on the entire distribution of $v^{\ast} - v^{\ast\ast}$, under the assumption that the value function of~\eqref{eq: problem} is Lipschitz continuous. 
This way, we can show that more and more probability mass of the duality gap distribution lies close to zero when $\card{\mathcal{C}}$ increases, demonstrating a ``left shift'' of the distribution. 
Finally, \Cref{theorem: probability lipschitz} bounds from below the probability that the value function is Lipschitz continuous and shows that this probability, too, approaches 1 as $\card{\mathcal{C}}$ increases, so that we can expect the Lipschitz assumption to hold in a ``nearly convex'' optimization problem.
 
Although the inherent randomness in electricity auctions causes the welfare maximization problem to vary daily, the i.i.d. assumption is clearly not satisfied in reality. Nonetheless, our theoretical results can help us understand why zero or vanishingly small duality gaps might frequently occur in partially nonconvex problems, as is the case empirically in electricity auctions. In this way, our study complements the worst-case bounds of \citet{aubin1976estimates} and brings us a step closer to understanding duality gaps.

\subsection{Example}\label{sec: example}

Let us consider a market with a fixed total demand of 5. The market is populated by a set of suppliers $\mathcal{I}$; a subset $\mathcal{C} \subseteq \mathcal{I}$ consists of continuous suppliers who can produce any quantity in the interval $[0,2]$, while the remaining suppliers in $\mathcal{N} = \mathcal{I} \setminus \mathcal{C}$ are discrete and can only produce quantities in the binary set $\{0,2\}$. The marginal costs $c_i$, $i \in \mathcal{I}$, are i.i.d.\ and continuously distributed, so that ties occur with probability zero. The corresponding welfare maximization problem is a mixed-integer linear program (MILP) formulated as:
\begin{subequations}\label{eq: knapsack}
    \begin{align}
        \min_x & \quad \sum_{i\in\mathcal{I}} c_i x_i \\
        \text{s.t.} & \quad \sum_{i\in\mathcal{I}} x_i = 5 \\
        & \quad x_i \in [0,2] \quad \forall i\in\mathcal{C} \\
        & \quad x_i \in \{0,2\} \quad \forall i\in\mathcal{N} .
    \end{align}
\end{subequations}
This setup is a special case of~\eqref{eq: problem} where the randomness is restricted entirely to the costs $c_i$. 

The problem structure is visualized in~\Cref{fig:random market}, where the vertical line represents the fixed demand of 5 and the suppliers are ordered along the horizontal axis from lowest to highest marginal cost. Suppliers with discrete production ($\mathcal{N}$) are illustrated by dashed horizontal steps, whereas suppliers with continuous capacities ($\mathcal{C}$) are shown as solid lines.

\begin{figure}[t]
\centering
\begin{tikzpicture}
    \begin{axis}[
        width=8cm, height=6cm,
        xmin=-0.5, xmax=10.5,
        ymin=-0.2, ymax=8.5,
        xtick={0,1,2,3,4,5,6,7,8,9,10},
        ytick={0,1,2,3,4,5,6,7,8},
        grid=both,
        grid style={black!10, thin}, 
        axis lines=left,
        axis line style={black, thick},
        tick label style={font=\small, black},
        xlabel={Quantity},
        x label style={at={(axis description cs:0.5,-0.1)}, anchor=north, black, font=\small},
        ylabel={Price},
        y label style={at={(axis description cs:-0.06,0.5)}, anchor=south, rotate=0, black, font=\small}
    ]
        % --- STAIRCASE FUNCTION PATHS ---
        \draw[cbDarkNavy, line width=1.8pt] (axis cs:0,1) -- (axis cs:2,1);
        \draw[cbDarkNavy, line width=1.8pt] (axis cs:2,1) -- (axis cs:2,3);
        \draw[cbDarkNavy, line width=1.8pt, dotted, dash pattern=on 4pt off 4pt] (axis cs:2,3) -- (axis cs:4,3);
        \draw[cbDarkNavy, line width=1.8pt] (axis cs:4,3) -- (axis cs:4,4);
        \draw[cbDarkNavy, line width=1.8pt] (axis cs:4,4) -- (axis cs:6,4);
        \draw[cbDarkNavy, line width=1.8pt] (axis cs:6,4) -- (axis cs:6,6);
        \draw[cbDarkNavy, line width=1.8pt, dotted, dash pattern=on 4pt off 4pt] (axis cs:6,6) -- (axis cs:8,6);
        \draw[cbDarkNavy, line width=1.8pt] (axis cs:8,6) -- (axis cs:8,7);
        \draw[cbDarkNavy, line width=1.8pt] (axis cs:8,7) -- (axis cs:10,7);
        % --- DATA NODE MARKERS ---
        \addplot[only marks, mark=*, mark size=2.8pt, fill=cbDarkNavy, draw=cbDarkNavy] coordinates {
            (0,1) (2,1) (2,3) (4,3) (4,4) (6,4) (6,6) (8,6) (8,7) (10,7)
        };
        % --- VERTICAL SEPARATOR LINE AT X=5 ---
        \draw[black, line width=1.8pt] (axis cs:5,0) -- (axis cs:5,8);
    \end{axis}
\end{tikzpicture}
\caption{Illustration of a simple welfare maximization problem.}\label{fig:random market}
\end{figure}

The duality gap of this mixed-integer program is exactly zero if and only if the third cheapest supplier belongs to the continuous set $\mathcal{C}$. 
This is because, in this case, the linear programming (LP) relaxation yields the same optimal value as the MILP.
Similarly, a Walrasian equilibrium exists whenever the segment for the third cheapest supplier in \Cref{fig:random market} is solid rather than dashed, since in that case the supply and demand curves intersect at a market-clearing price.
Because the costs $c_i$ are i.i.d., each supplier has an equal probability of occupying any position in the cost ranking. Consequently, the probability that the marginal supplier is continuous is exactly equal to the ratio of continuous suppliers in the market, given by $\card{\mathcal{C}}/\card{\mathcal{I}}$. 
Hence, the probability of a zero duality gap and equilibrium existence is $\card{\mathcal{C}}/\card{\mathcal{I}}$. 
If the number of discrete suppliers $\card{\mathcal{N}}$ remains fixed while the number of continuous suppliers $\card{\mathcal{C}}$ grows, the probability of obtaining a zero duality gap approaches 1.

\subsection{Literature Review}\label{subsec: literature review}

Our work contributes to distinct but intersecting streams of literature in mathematical optimization, theoretical economics, and electricity market design.

First, our study is related to the foundational work of \citet{aubin1976estimates} and subsequent articles such as \citet{bertsekas1982estimates}, \citet{udell2016bounding}, \citet{wang2017vanishing}, \citet{bi2020duality}, \citet{kerdreux2023stable}, and \citet{dey2026asymptotically}, which develop bounds on the duality gap of separable problems of type~\eqref{eq: problem} using the Shapley--Folkman lemma. 
We use the lemma and the argument behind it in a similar manner, but focus on partially nonconvex and random problems.

Second, our study is related to the literature initiated by \citet{dyer1989probabilistic} and continued by works like \citet{dey2021branch}, \citet{borst2023integrality}, and \citet{celaya2023proximity}, which analyze integer linear programs under i.i.d. or other stochastic assumptions on the underlying data. These papers typically focus on pure integer programs rather than mixed-integer programs. Our analysis differs because we explicitly consider partially nonconvex problems of the form~\eqref{eq: problem}, which are a generalization of mixed-integer linear programs.

Third, our work connects to a body of literature on the existence of Walrasian equilibria in nonconvex markets. \citet{starr1969quasi} originally introduced the Shapley--Folkman lemma to study approximate equilibrium existence in nonconvex markets, prior to its use in optimization by \citet{aubin1976estimates}. This connection is natural: a Walrasian equilibrium exists in a market if and only if the market's welfare maximization problem has a zero duality gap. Related works such as \citet{kelso1982job,bikhchandani1997competitive,baldwin2019understanding}, and \citet{tran2019product} study equilibrium existence when the traded commodities are indivisible (i.e., the welfare maximization problem is an integer program). We complement these contributions by examining the distribution of approximate equilibrium existence in a random market populated by a mix of agents with convex and nonconvex preferences.

Fourth, our study relates to research on nonconvex electricity auctions. Day-ahead electricity auctions feature nonconvex bids, which implies that a Walrasian equilibrium may not exist. The clearing algorithm EUPHEMIA is designed to detect a Walrasian equilibrium if one exists, and otherwise approximate it according to predefined rules by solving a variant of the underlying welfare maximization problem~\citep{euphemia2025}. Research in this domain generally focuses on designing alternative mechanisms to approximate a Walrasian equilibrium while balancing several conflicting objectives. Examples of such works are \citet{o2005efficient}, \citet{liberopoulos2016critical}, \citet{bichler2023pricing}, \citet{stevens2024some}, \citet{ahunbay2025pricing}, \citet{jungo2026comparing}, and \citet{milgrom2026walrasian}. 

We contribute to the literature on electricity auctions by analyzing commercially available data from the European day-ahead electricity auction, thereby providing empirical insights into the existence of an equilibrium. Although some related empirical findings are reported in annual reports of EUPHEMIA~\citep{cacm_report_2023}, they offer only limited information on duality gaps and equilibrium existence. In contrast, our results provide explicit upper bounds on the duality gaps of the welfare maximization problem across 281 days and nine countries. Ultimately, our theoretical framework helps explain the distribution of duality gaps observed in this important real-world optimization problem.

\section{Duality Gap}\label{sec: duality gap}

In this section, we derive lower bounds on the probability distribution of the duality gap $v^\ast - v^{\ast\ast}$ of problem~\eqref{eq: problem}.
We start with assumptions and preliminaries in \Cref{subsec: assumptions,subsec: preliminaries}, study the likelihood of a zero duality gap in \Cref{subsec: zero duality gap}, and continue with the case of a nonzero duality gap in \Cref{subsec: nonzero duality gap,subsec: value function}. Finally, in \Cref{subsec: summary}, we provide a short summary of those results before we continue to the applied part in \Cref{sec: equilibrium,sec: electricity}.

\subsection{Assumptions}\label{subsec: assumptions}

Let $(\Omega,\mathcal{F},\Pr)$ be a complete probability space and let $f_i$, $i\in\mathcal{I}$, be random functions on it. That is, $f_i=f_i(\omega,\cdot)$ with $f_i(\omega,\cdot):\mathbb{R}^n\to(\mathbb{R}\cup\{\infty\})^{m+1}$ for every $\omega\in\Omega$, and $\omega\mapsto f_i(\omega,x)$ is measurable for every fixed $x\in\mathbb{R}^n$. For simplicity, we suppress the argument $\omega$ throughout. We denote by $\mathcal{F}(\mathbb{R}^{n+m+1})$ the closed subsets of $\mathbb{R}^{n+m+1}$, equipped with the Effros $\sigma$-algebra generated by the sets $\{F\in\mathcal{F}(\mathbb{R}^{n+m+1}) \mid F\cap G\ne\emptyset\}$ for open $G\subseteq\mathbb{R}^{n+m+1}$. A measurable map from $(\Omega,\mathcal{F})$ into $\mathcal{F}(\mathbb{R}^{n+m+1})$ is called a \textit{random closed set}~\citep{molchanov2017theory}. We make the following assumptions on the random functions $f_i$.

\needspace{6\baselineskip}
\begin{assumption}\label{assumption: random functions}
The following holds for almost every $\omega\in\Omega$ and all $i\in\mathcal{I}$.
\begin{enumerate}[label=(\alph*)]
    \item \emph{Regular}: 
    $f_i$ is lower semicontinuous and its effective domain $\dom{f_i} := \big\{ x_i\in\mathbb{R}^n \mid f_i^k(x_i) < \infty, \; \forall k\in\mathcal{K}_0 \big\}$ is nonempty and compact.
    Moreover, its image $\im{f_i}:=\{ f_i(x_i) \mid x_i\in\dom{f_i} \}$ is bounded by $\im{f_i}\subseteq[-\gamma,\gamma]^{m+1}$ with a deterministic constant $\gamma$.
    Finally, a feasible point $\bar{x} \in \prod_{i\in\mathcal{I}} \dom{f_i}$ with $\sum_{i\in\mathcal{I}} f_i^k(\bar{x}_i) \le 0$, $k\in\mathcal{K}$, exists. 
    \item \emph{Partially nonconvex}: For $i\in\mathcal{C}$ the epigraph $\epi{f_i}$ is convex; for $i\in\mathcal{N}$ it might be nonconvex. The sets $\mathcal{C}$ and $\mathcal{N}$ are deterministic.
    \item \emph{Measurable epigraph}: The map $\omega\mapsto\epi{f_i}$ is measurable, i.e., a random closed set.
    \item \emph{Independent and identically distributed}: The random closed sets $\co{\epi{f_i}}$, $i\in\mathcal{I}$, are independent and identically distributed (i.i.d.).
\end{enumerate}
\end{assumption}

Part~(a) is a standard regularity condition in variational analysis and guarantees that the infima are attained (the bounded image assumption is not necessary for this, but we will need it later).
Part~(a) also makes part~(c) well posed: compactness of $\dom{f_i}$ together with lower semicontinuity implies that $\epi{f_i}$ is closed, and that its convex hull $\co{\epi{f_i}}$ is closed as well. Closedness alone, however, does not give measurability, and neither does pointwise measurability of $f_i$: the event $\{\omega \mid \epi{f_i}\cap G\ne\emptyset\}$ is an uncountable union of events indexed by $x$ and may fail to be measurable. We therefore require measurability in part~(c).

The convex hull $\co{\epi{f_i}}$ is then again a random closed set, because $F\mapsto\cl{\co{F}}$ is a measurable operation~\citep{molchanov2017theory}. Together with the completeness of $(\Omega,\mathcal{F},\Pr)$, this ensures that the random objects appearing in our analysis are measurable; measurability might otherwise fail because these are obtained by projecting $\epi{f_i}$, and the projection of a measurable set need not be measurable~\citep{molchanov2017theory}.

Part~(d) puts the i.i.d.\ assumption on the random sets $\co{\epi{f_i}}$, and no distributional assumption on $f_i$ or $\epi{f_i}$ themselves. By part~(b) we have $\co{\epi{f_i}}=\epi{f_i}$ for $i\in\mathcal{C}$, so the family $(\epi{f_i})_{i\in\mathcal{C}}$, and hence $(f_i)_{i\in\mathcal{C}}$, is i.i.d.\ as well; for $i\in\mathcal{N}$ this need not be the case.

\subsection{Preliminaries}\label{subsec: preliminaries}

As our analysis relies mainly on geometric arguments, we introduce geometric versions of~\eqref{eq: problem} and~\eqref{eq: dual} that are easier to work with. For this, we use the projection of $\epi{f_i}$ on the value-space:
\begin{equation*}
    V_i := \big\{ z_i\in\mathbb{R}^{m+1} \mid \exists x_i\in\mathbb{R}^n \text{ s.t. } z_i \ge f_i(x_i) \big\}
\end{equation*}
where $z_i=(z_i^k)_{k\in\mathcal{K}_0}$. Equivalently, $V_i=\im{f_i}+\mathbb{R}^{m+1}_{\ge0}$, because $z_i\ge f_i(x_i)$ requires $x_i\in\dom{f_i}$.
It is easy to see that problem~\eqref{eq: problem} can be equivalently stated as
\begin{subequations}\label{eq: primal epigraph}
    \begin{align}
        v^\ast = \;\; \inf_z & \quad \sum_{i\in\mathcal{I}} z_i^0 \\
        \text{s.t.} & \quad \sum_{i\in\mathcal{I}} z_i^k \le 0 \quad \forall k\in\mathcal{K} \\
        & \quad z_i \in V_i \quad \forall i\in\mathcal{I} .
    \end{align}
\end{subequations}
Less obvious is that also the dual~\eqref{eq: dual} can be stated in such a form. This can be achieved by using the convex hull $\co{V_i}$ instead of $V_i$: 
\begin{subequations}\label{eq: dual epigraph}
    \begin{align}
        v^{\ast\ast} = \;\; \inf_z & \quad \sum_{i\in\mathcal{I}} z_i^0 \\
        \text{s.t.} & \quad \sum_{i\in\mathcal{I}} z_i^k \le 0 \quad \forall k\in\mathcal{K}  \\
        & \quad z_i \in \co{V_i} \quad \forall i\in\mathcal{I} .
    \end{align}
\end{subequations}

\begin{lemma}\label{lemma: equal value}
Problems~\eqref{eq: dual} and~\eqref{eq: dual epigraph} have the same optimal value $v^{\ast\ast}$.
\end{lemma}
\begin{proof}
    Compactness of $\dom{f_i}$ and lower semicontinuity of $f_i$ imply that $V_i$ is closed, and $\co{V_i}=\co{\cl{\im{f_i}}}+\mathbb{R}^{m+1}_{\ge0}$ is closed as the sum of a compact and a closed set. The statement therefore follows from \citet[Theorem~2.3]{lemarechal2001geometric}.
\end{proof}

From \Cref{assumption: random functions}(a), we can deduce that the infima in~\eqref{eq: primal epigraph} and~\eqref{eq: dual epigraph} are finite and attained.

\begin{lemma}\label{lemma: attainment}
There is an optimal solution to~\eqref{eq: primal epigraph} and~\eqref{eq: dual epigraph} with $-\infty < v^{\ast\ast} \le v^\ast < \infty$.
\end{lemma}
\begin{proof}
    By \Cref{assumption: random functions}(a), $z_i=f_i(\bar{x}_i)$ is feasible for~\eqref{eq: primal epigraph} and, since $V_i\subseteq\co{V_i}$, also for~\eqref{eq: dual epigraph}; as every $z_i^0\ge-\gamma$, this gives $-\card{\mathcal{I}}\gamma\le v^{\ast\ast}\le v^\ast<\infty$. 
    
    For attainment, restrict the feasible set of either problem to $\{z \mid \sum_{i\in\mathcal{I}} z_i^0\le v^\ast\}$, which leaves the optimal value unchanged. On this set, every coordinate is bounded above: since all $z_i^k\ge-\gamma$, isolating one index in the constraint $\sum_{i\in\mathcal{I}} z_i^k\le0$ gives $z_j^k\le(\card{\mathcal{I}}-1)\gamma$ for $k\in\mathcal{K}$, and likewise
    $z_j^0\le v^\ast+(\card{\mathcal{I}}-1)\gamma$. The restricted set is therefore compact, and the continuous objective attains its minimum.
\end{proof}

Now we know that the duality gap $v^\ast-v^{\ast\ast}$ exists, and we can study it through studying problems~\eqref{eq: primal epigraph} and~\eqref{eq: dual epigraph}.
Finally, we confirm that the i.i.d. property of $\co{\epi{f_i}}$ transfers to $\co{V_i}$.

\begin{lemma}\label{lemma: iid projections}
    The random closed sets $\co{V_i}$, $i\in\mathcal{I}$, are i.i.d.
\end{lemma}
\begin{proof}
    Let $\pi:\mathbb{R}^{n+m+1}\to\mathbb{R}^{m+1}$, $\pi(x,z)=z$, be the projection on the value-space. Since $\pi$ is linear, it commutes with taking convex hulls, so $\pi(\co{\epi{f_i}})=\co{\pi(\epi{f_i})}=\co{V_i}$, which is closed (\Cref{lemma: equal value}). As $\pi^{-1}(G)$ is open for every open $G\subseteq\mathbb{R}^{m+1}$, the events $\{\co{V_i}\cap G\ne\emptyset\}$ and $\{\co{\epi{f_i}}\cap\pi^{-1}(G)\ne\emptyset\}$ coincide, so $\co{V_i}$ is a random closed set obtained from $\co{\epi{f_i}}$ by a deterministic measurable
    map. Applying the same map to the i.i.d.\ sets $\co{\epi{f_i}}$ therefore preserves independence and equality of laws~\citep{molchanov2017theory}.
\end{proof}

\subsection{Zero Duality Gap}\label{subsec: zero duality gap}

We denote by $z^\ast=(z^\ast_i)_{i\in\mathcal{I}}$ a solution to~\eqref{eq: primal epigraph}. Similarly, we denote by $z^{\ast\ast}$ a possible solution to~\eqref{eq: dual epigraph}. We are interested in the proximity of $z^\ast$ and $z^{\ast\ast}$.
This proximity can be characterized by the number of indices $i\in\mathcal{I}$ for which $z^{\ast\ast}_i\notin V_i$. If this number is zero, then $z^{\ast\ast}$ is also a solution to~\eqref{eq: primal epigraph} and the duality gap $v^\ast-v^{\ast\ast}$ must be zero. A larger number of such indices indicates a greater distance between the dual and primal solutions. Consequently, we would like to bound this number.
For this, \citet{aubin1976estimates} pioneered the use of the Shapley--Folkman lemma~\citep{starr1969quasi}, which we state and use in an alternative proof in~\Cref{subsec: convexified market}. Here, we do not use the lemma itself but the dimensional argument behind it: since $m+2$ vectors in $\mathbb{R}^{m+1}$ are linearly dependent, a solution to~\eqref{eq: dual epigraph} can be chosen such that at most $m+1$ of its points $z_i$ are not extreme points of $\co{V_i}$.

\begin{lemma}\label{lemma: extremepoints}
There exist a solution $z'$ to~\eqref{eq: dual epigraph} and an index set $\mathcal{S} \subseteq \mathcal{I}$ with $\card{\mathcal{S}} = \card{\mathcal{I}} - m - 1$ such that
\begin{equation}\label{eq: extremepoint}
     z_i' \in \ext{\co{V_i}} \quad \forall i\in\mathcal{S} .
\end{equation}
In particular, $z_i'\in V_i$ for all $i\in\mathcal{S}$.
\end{lemma}
\begin{proof}
The set of solutions to~\eqref{eq: dual epigraph} is nonempty by \Cref{lemma: attainment}, closed, and convex. It contains no line, because $\co{V_i}\subseteq[-\gamma,\infty)^{m+1}$ for all $i\in\mathcal{I}$. Hence it has an extreme point $z'$~\citep[Corollary~18.5.3]{rockafellar1970convex}.

Suppose that $z_i'\notin\ext{\co{V_i}}$ for $m+2$ indices $i$. For each of them, there is a vector $e_i\neq0$ with $z_i'\pm e_i\in\co{V_i}$. Since $m+2$ vectors in $\mathbb{R}^{m+1}$ are linearly dependent, there are scalars $t_i$, not all zero, with $\lvert t_i\rvert\le1$ and $\sum_i t_ie_i=0$. Replacing $z_i'$ by $z_i'+t_ie_i$, respectively by $z_i'-t_ie_i$, for these indices keeps every $z_i'$ in $\co{V_i}$ and leaves all sums $\sum_{i\in\mathcal{I}}z_i'$ unchanged. Both modified collections are therefore solutions to~\eqref{eq: dual epigraph}, and $z'$ is their midpoint, which contradicts the extremality of $z'$. Hence $z_i'\notin\ext{\co{V_i}}$ holds for at most $m+1$ indices, and any $\card{\mathcal{I}}-m-1$ of the remaining indices form the set $\mathcal{S}$. Finally, every extreme point of $\co{V_i}$ lies in $V_i$, because a point of $\co{V_i}$ outside $V_i$ is a nontrivial convex combination of points of $V_i$.
\end{proof}

\Cref{lemma: extremepoints} tells us that there is a set $\mathcal{S}$ consisting of $\card{\mathcal{I}} - m - 1$ indices $i\in\mathcal{I}$ with $z_i' \in \ext{\co{V_i}}\subseteq V_i$. To understand which indices $i$ lie in $\mathcal{S}$, we can characterize the distribution of $\mathcal{S}$.

\begin{lemma}\label{lemma: uniform distribution}
    Let $\mathcal{S}$ be the index set of \Cref{lemma: extremepoints}, chosen uniformly at random whenever there are several. Then every subset of $\mathcal{I}$ with cardinality $\card{\mathcal{I}}-m-1$ is equally likely to be $\mathcal{S}$.
\end{lemma}
\begin{proof}
    Let $\sigma$ be a permutation of $\mathcal{I}$ and consider the relabeled family $(\co{V_{\sigma(i)}})_{i \in \mathcal{I}}$.
    
    Whether a given index set satisfies~\eqref{eq: extremepoint} for some solution to~\eqref{eq: dual epigraph} depends only on $(\co{V_i})_{i \in \mathcal{I}}$ itself, not on the labels of the indices: problem~\eqref{eq: dual epigraph} takes only $(\co{V_i})_{i \in \mathcal{I}}$ as input, and condition~\eqref{eq: extremepoint} refers only to the extreme points of $\co{V_i}$. The tie-breaking rule (``chosen uniformly at random whenever there are several'') treats all indices symmetrically as well.
    Consequently, if $\mathcal{S}$ is the set selected for the original input $(\co{V_i})_{i \in \mathcal{I}}$, then $\sigma^{-1}(\mathcal{S})$ is the set selected for the relabeled input $(\co{V_{\sigma(i)}})_{i \in \mathcal{I}}$.
    
    Since the $\co{V_i}$ are i.i.d.\ (\Cref{lemma: iid projections}), the relabeled family $(\co{V_{\sigma(i)}})_{i \in \mathcal{I}}$ has the same distribution as $(\co{V_i})_{i \in \mathcal{I}}$, and hence $\sigma^{-1}(\mathcal{S})$ has the same distribution as $\mathcal{S}$. Thus, for every subset $\mathcal{L}\subseteq\mathcal{I}$ with cardinality $\card{\mathcal{I}}-m-1$,
    \begin{equation*}
        \Pr(\mathcal{S} = \mathcal{L})
        \; = \; \Pr\big(\sigma^{-1}(\mathcal{S}) = \mathcal{L}\big)
        \; = \; \Pr\big(\mathcal{S} = \sigma(\mathcal{L})\big) .
    \end{equation*}
    Since any such subset can be mapped to any other by a suitable permutation $\sigma$, all of them are equally likely.
\end{proof}

We can use \Cref{lemma: extremepoints,lemma: uniform distribution} to say something about the probability of a zero duality gap. The duality gap vanishes whenever the set $\mathcal{S}$ contains all nonconvex indices, because $V_i=\co{V_i}$ holds for the remaining, convex ones.
Because $\mathcal{S}$ is a uniformly random subset of $\mathcal{I}$, the number of nonconvex indices it contains follows a hypergeometric distribution. Evaluating its probability mass function at $\card{\mathcal{N}}$ then gives the lower bound.

\begin{theorem}\label{theorem: zero duality gap}
The probability of a zero duality gap is bounded below by:
\begin{equation*}
    \Pr\big[v^\ast-v^{\ast\ast}=0\big] \ \ge \
    \frac{\binom{\card{\mathcal{C}}}{m+1}}{\binom{\card{\mathcal{I}}}{m+1}} ,
\end{equation*}
where $\binom{n}{k} = \frac{n!}{k!(n-k)!}$ denotes the binomial coefficient.
\end{theorem}
\begin{proof}
Let $\mathcal{S}$ be the index set of \Cref{lemma: extremepoints}, chosen uniformly at random whenever there are several, and let $z'$ be a corresponding solution to~\eqref{eq: dual epigraph}. By \Cref{lemma: uniform distribution}, $\mathcal{S}$ is uniformly distributed over all subsets of $\mathcal{I}$ with cardinality $\card{\mathcal{I}}-m-1$.

Let $\mathcal{S}'=\mathcal{N}\cap\mathcal{S}$ collect the nonconvex indices in $\mathcal{S}$, and suppose that $\card{\mathcal{S}'}=\card{\mathcal{N}}$, i.e., that $\mathcal{S}$ contains all of them. The corresponding solution $z'$ then satisfies $z'_i\in\ext{\co{V_i}}\subseteq V_i$ for all $i\in\mathcal{N}$, and $z'_i\in\co{V_i}=V_i$ for all $i\in\mathcal{C}$. Hence $z'$ is feasible for~\eqref{eq: primal epigraph} and attains the objective value $v^{\ast\ast}$ there, so that $v^\ast-v^{\ast\ast}=0$. It therefore suffices to evaluate the probability of the event $\card{\mathcal{S}'}=\card{\mathcal{N}}$. 

Because $\mathcal{S}$ is uniformly distributed over all subsets of $\mathcal{I}$ with cardinality $\card{\mathcal{I}}-m-1$, drawing the set $\mathcal{S}$ amounts to drawing $\card{\mathcal{I}}-m-1$ indices without replacement from the $\card{\mathcal{I}}$ indices, of which $\card{\mathcal{N}}$ are nonconvex. Hence, $\card{\mathcal{S}'}$ follows a hypergeometric distribution: $$\card{\mathcal{S}'}\sim\mathrm{Hypergeom}(\card{\mathcal{I}},\card{\mathcal{N}}, \card{\mathcal{I}}-m-1).$$ Evaluating its probability mass function at $\card{\mathcal{S}'}=\card{\mathcal{N}}$ and using the symmetry $\binom{n}{k}=\binom{n}{n-k}$ gives
\begin{equation*}
    \Pr(\card{\mathcal{S}'} = \card{\mathcal{N}}) \; = \;
    \frac{\binom{\card{\mathcal{C}}}{\card{\mathcal{C}}-m-1}}
    {\binom{\card{\mathcal{I}}}{\card{\mathcal{I}}-m-1}} \; = \;
    \frac{\binom{\card{\mathcal{C}}}{m+1}}{\binom{\card{\mathcal{I}}}{m+1}} ,
\end{equation*}
which yields the claimed lower bound.
\end{proof}

\tikzset{
    declare function={
        binom(\n,\k) = \n! / (\k! * (\n-\k)!);
        ratioFunc(\C,\I,\m) = binom(\C, \m+1) / binom(\I, \m+1);
    }
}

\begin{figure}[t]
\centering
\begin{tikzpicture}
    \begin{axis}[
        width=11cm, height=7.5cm,
        xmin=0, xmax=52,
        ymin=-0.05, ymax=1.05,
        xtick={0, 10, 20, 30, 40, 50},
        ytick={0, 0.2, 0.4, 0.6, 0.8, 1.0},
        grid=both, 
        grid style={cbMidnightBlue!10},
        axis lines=middle, 
        inner axis line style={-Stealth},
        tick label style={font=\small},
        xlabel={Number of convex functions $\card{\mathcal{C}}$ for fixed $\card{\mathcal{I}}=50$},
        x label style={at={(axis description cs:0.5,-0.1)}, anchor=north, font=\small},
        ylabel={Probability},
        y label style={at={(axis description cs:-0.08,0.5)}, anchor=south, rotate=90, font=\small},
        legend style={at={(axis description cs:0.05,0.92)}, anchor=north west, font=\footnotesize}
    ]
        % Curve 1: m = 1 (Combinations of size 2)
        \addplot[
            domain=2:50, samples=49, 
            dotted,
            ultra thick, black
        ] {ratioFunc(x, 50, 1)};
        \addlegendentry{$m = 1$};
        % Curve 2: m = 3 (Combinations of size 4)
        \addplot[
            domain=4:50, samples=47, 
            dashed,
            ultra thick, black
        ] {ratioFunc(x, 50, 3)};
        \addlegendentry{$m = 3$};
        % Curve 3: m = 5 (Combinations of size 6)
        \addplot[
            domain=6:50, samples=45, 
            ultra thick, black
        ] {ratioFunc(x, 50, 5)};
        \addlegendentry{$m = 5$};
    \end{axis}
\end{tikzpicture}
\caption{The ratio $\frac{\binom{\card{\mathcal{C}}}{m+1}}{\binom{\card{\mathcal{I}}}{m+1}}$ for a fixed $\card{\mathcal{I}}=50$.}\label{fig:combinatorial_ratio}
\end{figure}

\Cref{fig:combinatorial_ratio} plots the bound of \Cref{theorem: zero duality gap} for a fixed number of functions $\card{\mathcal{I}}=50$ and several numbers of constraints $m$. The bound increases monotonically in the number of convex functions $\card{\mathcal{C}}$ and equals 1 once all functions are convex, so a zero duality gap becomes more likely the ``more convex'' the problem is. The bound is weaker the more constraints the problem has, as \Cref{lemma: extremepoints} then leaves more indices unaccounted for.

\subsection{Nonzero Duality Gap}\label{subsec: nonzero duality gap}

To extend \Cref{theorem: zero duality gap} to the nonzero duality gap case $v^\ast-v^{\ast\ast}>0$, we need to measure the nonconvexity of $f_i$. For this, we use the directed Hausdorff distance from the convex hull $\co{V_i}$ of the epigraph-projection to the epigraph-projection $V_i$ itself:
\begin{equation*}
    \alpha(f_i) \; := \; \sup_{z \in \co{V_i}} \; \inf_{y \in V_i} \; \norm{z - y}_{\infty} .
\end{equation*}
For all $i\in\mathcal{C}$ we have that $\alpha(f_i)=0$. 
The maximal nonconvexity is given by
\begin{equation*}
    \alpha \; := \; \max_{i\in\mathcal{I}} \; \alpha(f_i) .
\end{equation*}
Note that $\alpha$ is random as the functions $f_i$ are random.
However, by the bounded image $\im{f_i}\subseteq[-\gamma,\gamma]^{m+1}$ in \Cref{assumption: random functions}(a), we can bound $\alpha$ deterministically using $\gamma$. 

\begin{lemma}
It holds that $\alpha \; \le \; 2\gamma$.
\end{lemma}
\begin{proof}
Recall from \Cref{subsec: preliminaries} that $V_i = \im{f_i} + \mathbb{R}^{m+1}_{\ge 0}$, which implies that its convex hull is $\co{V_i} = \co{\im{f_i}} + \mathbb{R}^{m+1}_{\ge 0}$. 

Any point $z \in \co{V_i}$ can be decomposed as $z = z_{\text{co}} + d_z$, where $z_{\text{co}} \in \co{\im{f_i}}$ and $d_z \in \mathbb{R}^{m+1}_{\ge 0}$. For a fixed $z$ and any arbitrary choice of $w \in \im{f_i}$, the point $y^* = w + d_z$ lies in $V_i$. Evaluating the directed distance yields:
\begin{equation*}
    \min_{y \in V_i} \norm{z - y}_{\infty} \; \le \; \norm{z - y^*}_{\infty} \; = \; \norm{(z_{\text{co}} + d_z) - (w + d_z)}_{\infty} \; = \; \norm{z_{\text{co}} - w}_{\infty} .
\end{equation*}
Because $\im{f_i} \subseteq [-\gamma, \gamma]^{m+1}$ it holds that $\co{\im{f_i}} \subseteq [-\gamma, \gamma]^{m+1}$. Hence, both $z_{\text{co}}$ and $w$ lie within $[-\gamma, \gamma]^{m+1}$ and the statement follows.
\end{proof}

Let us now consider the value function of problem~\eqref{eq: problem}:
\begin{subequations}\label{eq: value problem}
    \begin{align}
        v(b) \; := \; \min_z & \quad \sum_{i\in\mathcal{I}} z_i^0  \\
        \text{s.t.} & \quad \sum_{i\in\mathcal{I}} z_i^k \le b^k \quad \forall k\in\mathcal{K} \\
        & \quad z_i \in V_i \quad \forall i\in\mathcal{I} ,
    \end{align}
\end{subequations}
where $b=(b^k)_{k\in\mathcal{K}}$.
The value function $v(b)$ measures the sensitivity of the optimal value to changes in the constraints. Note that $v(0)=v^\ast$. The domain $\dom{v}\subseteq\mathbb{R}^m$ equals all the points where the problem is feasible. 
The feasible point in \Cref{assumption: random functions}(a) implies that $\mathbb{R}_{\ge 0}^m \subseteq \dom{v}$.

For convex optimization problems, the value function $v(b)$ is known to be Lipschitz continuous under weak regularity assumptions~\citep{BoydVandenberghe2004}. 
In contrast, for nonconvex optimization problems, it might be discontinuous~\citep{blair1977value,blair1979value}.
For the partially nonconvex case, we can argue that the likelihood that the value function is Lipschitz continuous increases with the number of convex functions relative to the number of nonconvex functions. We will do so in \Cref{subsec: value function}. 
For now, let us assume that the value function is Lipschitz continuous over the positive box 
\begin{equation}\label{eq: box}
    \mathcal{B} \; = \; \big\{ b\in\mathbb{R}_{\ge0}^m \ \big| \ \norm{b}_{\infty} \le (m+1) \cdot 2\gamma \big\} .
\end{equation}
Here and throughout, Lipschitz continuity with constant $L$ is understood with respect to the $\infty$-norm, i.e., $\lvert v(b_1)-v(b_2)\rvert\le L\norm{b_1-b_2}_\infty$.

With this assumption in place, we can extend \Cref{theorem: zero duality gap} to the nonzero duality gap
case, by arguing as follows:
We start from \Cref{lemma: extremepoints} again and use that $z_i'\notin V_i$ for at most $m+1$ indices $i$. When shifting these points to nearby points of $V_i$, we know from $\alpha\le2\gamma$ that this shift is bounded. Since $v(b)$ is Lipschitz over this bounded area, we know that the loss in objective value must be bounded by a constant proportional to $m+1$. Formally:

\begin{theorem}\label{theorem: non-zero duality gap}
Let $v(b)$ be Lipschitz continuous over $\mathcal{B}$ with constant $L$. 
Then
\begin{equation*}
    \Pr\big[v^{\ast} - v^{\ast\ast} \le (L+1) 2\gamma \cdot t\big] \ \ge \ \sum_{r=0}^{t} \frac{\binom{\card{\mathcal{N}}}{r}\binom{\card{\mathcal{C}}}{m+1-r}}{\binom{\card{\mathcal{I}}}{m+1}}
\end{equation*}
for $t=0,1,\ldots,m+1$.
\end{theorem}
\begin{proof}
Let $z'$ be the solution to~\eqref{eq: dual epigraph} guaranteed by \Cref{lemma: extremepoints}, which satisfies $z'_i \in \ext{\co{V_i}}\subseteq V_i$ for all $i \in \mathcal{S}$ where $\card{\mathcal{S}} = \card{\mathcal{I}} - m - 1$. Let $\mathcal{J} = \mathcal{I} \setminus \mathcal{S}$ denote the remaining set of indices, so that $\card{\mathcal{J}} = m + 1$. Let us define $\mathcal{J}_{\mathcal{N}} = \mathcal{J} \cap \mathcal{N}$ as the set of nonconvex indices excluded from $\mathcal{S}$, and let $r = \card{\mathcal{J}_{\mathcal{N}}}$.
By definition of $\mathcal{S}$ and the fact that $\co{V_i} = V_i$ for all $i \in \mathcal{C}$, we know that $z'_i \in V_i$ for all $i \in \mathcal{I}\setminus\mathcal{J}_{\mathcal{N}}$. 

For $i \in \mathcal{J}_{\mathcal{N}}$, since $z'_i \in \co{V_i}$, the definition of $\alpha(f_i)$ guarantees a point $y_i \in V_i$ with $\norm{y_i - z'_i}_{\infty} \le \alpha$. For all $i \in \mathcal{I}\setminus\mathcal{J}_{\mathcal{N}}$, set $y_i = z'_i \in V_i$. By construction, $y_i - z_i' = 0$ for $i\notin\mathcal{J}_{\mathcal{N}}$ and $\norm{y_i-z_i'}_\infty\le\alpha$ for $i\in\mathcal{J}_{\mathcal{N}}$. Hence, for every $k\in\mathcal{K}_0$,
\begin{subequations}
\begin{equation}\label{eq: uniform shift bound}
    \sum_{i\in\mathcal{I}} y_i^k \;-\; \sum_{i\in\mathcal{I}} z_i^{\prime,k} \;=\; \sum_{i\in\mathcal{J}_{\mathcal{N}}} \big(y_i^k - z_i^{\prime,k}\big) \;\le\; r\alpha .
\end{equation}

Applying~\eqref{eq: uniform shift bound} to $k\in\mathcal{K}$ and using feasibility of $z'$ for~\eqref{eq: dual epigraph} (i.e., $\sum_{i\in\mathcal{I}} z_i^{\prime,k}\le0$), we get

\begin{equation}\label{eq: constraint bound}
    \sum_{i\in\mathcal{I}} y_i^k \;\le\; \sum_{i\in\mathcal{I}} z_i^{\prime,k} + r\alpha \;\le\; r\alpha \qquad \forall k\in\mathcal{K}.
\end{equation}
Applying~\eqref{eq: uniform shift bound} to $k=0$ and using $\sum_{i\in\mathcal{I}} z_i^{\prime,0} = v^{\ast\ast}$, we get
\begin{equation}\label{eq: objective bound}
    \sum_{i\in\mathcal{I}} y_i^0 \;\le\; v^{\ast\ast} + r\alpha .
\end{equation}
\end{subequations}

\eqref{eq: constraint bound} implies that the collection $(y_i)_{i\in\mathcal{I}}$ is a feasible solution to the perturbed primal problem~\eqref{eq: value problem} at $b = r\alpha \mathbf{1}_m$ where $\mathbf{1}_m$ is the all-ones-vector. By definition of the value function and~\eqref{eq: objective bound},
\begin{equation*}
    v(r\alpha \mathbf{1}_m) \;\le\; \sum_{i\in\mathcal{I}} y_i^0 \;\le\; v^{\ast\ast} + r\alpha .
\end{equation*}
Using that $v(b)$ is Lipschitz continuous over $\mathcal{B}$ with constant $L$, we bound the primal optimal value $v^\ast = v(0)$:
\begin{equation*}
    v^\ast - v^{\ast\ast} \;\le\; v(0) - v(r\alpha \mathbf{1}_m) + r\alpha \;\le\; L \norm{r\alpha \mathbf{1}_m}_{\infty} + r\alpha \;=\; (L+1)\alpha r \;\le\; (L+1)2\gamma r .
\end{equation*}

Finally, as in the proof of \Cref{theorem: zero duality gap}, the index set $\mathcal{J}$ is uniformly random among all size-$(m+1)$ subsets of $\mathcal{I}$, so $r = \card{\mathcal{J}\cap\mathcal{N}} \sim \mathrm{Hypergeom}(\card{\mathcal{I}},\,\card{\mathcal{N}},\,m+1)$. Since $r\le t$ implies $v^\ast - v^{\ast\ast} \le (L+1)2\gamma t$, summing the hypergeometric probabilities from $r=0$ to $t$ gives the claimed lower bound.
\end{proof}

\Cref{theorem: non-zero duality gap} establishes that the cumulative distribution function (CDF) of the duality gap $v^\ast-v^{\ast\ast}$ is lower-bounded at the $m+2$ discrete points $0, (L+1)2\gamma, 2(L+1)2\gamma, \ldots, (m+1)(L+1)2\gamma$ by the CDF of the hypergeometric distribution $\mathrm{Hypergeom}(\card{\mathcal{I}},\, \card{\mathcal{N}},\, m+1)$. 

\begin{figure}[t]
\centering
\begin{tikzpicture}
    \begin{axis}[
        width=11.5cm, height=8cm,
        xmin=-0.5, xmax=5.5,
        ymin=-0.05, ymax=1.15,
        % Custom domain ticks representing the scaled values
        xtick={0, 1, 2, 3, 4, 5},
        xticklabels={$0$, $T$, $2T$, $3T$, $4T$, $5T$},
        ytick={0, 0.2, 0.4, 0.6, 0.8, 1.0},
        yticklabels={},
        grid=both, 
        grid style={cbMidnightBlue!10},
        axis lines=box, 
        tick label style={font=\small},
        % Labels placed horizontally below and vertically left
        xlabel={$x$},
        x label style={at={(axis description cs:0.5,-0.06)}, anchor=north, font=\small},
        ylabel={Cumulative Probability},
        y label style={at={(axis description cs:-0.01,0.5)}, anchor=south, font=\small}
    ]
        % 1. Stepwise Lower Bound: Hypergeometric CDF (Stochastic Dominator)
        % Increasing at 1T, 2T, 3T, and hitting exactly 1.0 at 4T
        \addplot[
            draw=black, 
            ultra thick, 
            dashed,
            const plot,
            domain=0:5,
            samples at={0, 1, 2, 3, 4, 5}
        ] coordinates {
            (0, 0.20)
            (1, 0.45)
            (2, 0.70)
            (3, 0.90)
            (4, 1.00)
            (5, 1.00)
        };
        % Open circles to indicate the step discontinuities cleanly
        \addplot[only marks, mark=o, mark size=2.5pt, draw=black, thick] coordinates {
            (1, 0.20) (2, 0.45) (3, 0.70) (4, 0.90) (5, 1.00)
        };
        % Closed circles showing the value at the jump points
        \addplot[only marks, mark=*, mark size=2.5pt, fill=black, draw=black] coordinates {
            (0, 0.20) (1, 0.45) (2, 0.70) (3, 0.90) (4, 1.00) (5, 1.00)
        };
        % 2. True Duality Gap CDF curve: Monotonic, strictly above the bound, flattening exactly at 1.0
        \addplot[
            draw=black, 
            ultra thick, 
            smooth, 
            domain=0:5, 
            samples=100
        ] coordinates {
            (0, 0.35)
            (0.6, 0.42)
            (1, 0.58)
            (1.5, 0.68)
            (2, 0.82)
            (2.7, 0.94)
            (3.3, 0.98)
            (4, 1.00)
            (5, 1.00)
        };
    \end{axis}
\end{tikzpicture}
\caption{Illustration of the CDF of the duality gap $v^\ast - v^{\ast\ast}$ (\textit{solid line}) and the lower-bounding CDF of the hypergeometric distribution where $T=(L+1)2\gamma$ (\textit{dashed line}).}\label{fig:stochastic_dominance_cdf_corrected}
\end{figure}

Since any probability CDF is a monotonically nondecreasing function, and the CDF of the discrete distribution $\mathrm{Hypergeom}(\card{\mathcal{I}},\, \card{\mathcal{N}},\, m+1)$ is stepwise constant over a continuous domain, this bound can be extended globally (cf.~\Cref{fig:stochastic_dominance_cdf_corrected}). Specifically, for any interval $x \in [t, t+1)$, the monotonic property implies $\Pr(v^\ast - v^{\ast\ast} \le (L+1)2\gamma x) \ge \Pr(v^\ast - v^{\ast\ast} \le (L+1)2\gamma t)$. Consequently, for all $x \in \mathbb{R}_{\ge0}$, it holds that
\begin{equation}\label{eq: stochastic dominance}
    \Pr(v^\ast - v^{\ast\ast} \le (L+1)2\gamma \cdot x) \;\; \ge \;\; \sum_{r=0}^{\lfloor x \rfloor} \frac{\binom{\card{\mathcal{N}}}{r}\binom{\card{\mathcal{C}}}{m+1-r}}{\binom{\card{\mathcal{I}}}{m+1}}.
\end{equation}
The inequality in \eqref{eq: stochastic dominance} demonstrates a relation of \textit{first-order stochastic dominance}. Specifically, the scaled random variable $(L+1)2\gamma \cdot R$, where $R \sim \mathrm{Hypergeom}(\card{\mathcal{I}},\, \card{\mathcal{N}},\, m+1)$, first-order stochastically dominates the duality gap $v^\ast - v^{\ast\ast}$. From this dominance relation, it follows directly that the expected value of the duality gap is upper-bounded by that of the scaled hypergeometric variable dominating it. Formally:

\begin{corollary}\label{corollary: expected duality gap}
Let $v(b)$ be Lipschitz continuous over $\mathcal{B}$ with constant $L$. Then
\begin{equation*}
    \E{v^\ast-v^{\ast\ast}} \ \le \ (L+1) 2\gamma \cdot (m+1) \cdot \frac{\card{\mathcal{N}}}{\card{\mathcal{I}}}.
\end{equation*}
\end{corollary}
\begin{proof}
Since both sides of the stochastic dominance relation represent nonnegative random variables, the statement follows immediately by applying the expectation operator to the first-order stochastically dominating variable $(L+1)2\gamma \cdot R$, where $\E{R} = (m+1)\frac{\card{\mathcal{N}}}{\card{\mathcal{I}}}$.
\end{proof}

\Cref{corollary: expected duality gap} tells us that the expected duality gap goes to zero with increasing share of convex functions (and thus decreasing $\frac{\card{\mathcal{N}}}{\card{\mathcal{I}}}$).

\begin{remark}\label{remark: no lipschitz}
    The Lipschitz assumption is needed only because moving a point $z_i'\in\co{V_i}$ to a point $y_i\in V_i$ may violate the coupling constraints. This cannot happen if all domains $\dom{f_i}$ are convex and all constraint functions $f_i^k$, $k\in\mathcal{K}$, are convex, so that the nonconvexity lies entirely in the objective functions $f_i^0$. Writing $z_i'=\sum_j\theta_j f_i(x_j)+d$ with $x_j\in\dom{f_i}$, convex weights $\theta_j$, and $d\ge0$, the point $y_i:=f_i(\sum_j\theta_jx_j)\in V_i$ then satisfies $y_i^k\le z_i^{\prime,k}$ for all $k\in\mathcal{K}$ by convexity, and $y_i^0\le z_i^{\prime,0}+2\gamma$. The collection $y$ is therefore feasible for~\eqref{eq: primal epigraph}, and the proof of \Cref{theorem: non-zero duality gap} yields $v^\ast-v^{\ast\ast}\le2\gamma r$. Hence the bound of \Cref{theorem: non-zero duality gap} holds with $L=0$ and without any Lipschitz assumption.
\end{remark}

\subsection{Lipschitz Value Function}\label{subsec: value function}

Our argument in \Cref{theorem: non-zero duality gap} required a Lipschitz-continuous value function. 
Now we study the likelihood that the value function $v(b)$ of problem~\eqref{eq: problem} is Lipschitz over $\mathcal{B}$ defined in~\eqref{eq: box}.
We will argue that the value function of the convex subproblem of~\eqref{eq: problem} is Lipschitz and then show that if $\card{\mathcal{C}}$ is large enough, then also problem~\eqref{eq: problem} has a Lipschitz value function given some regularity conditions on $f_i$.

\begin{remark}
    The sufficient condition derived below requires the constraints of~\eqref{eq: problem} to be strictly satisfiable. It therefore fails whenever an equality is written as two opposite inequalities, in which case the results of this subsection remain valid but become vacuous. Conceptually, the approach extends to affine equality constraints: the convex functions must then be able to absorb deviations of the nonconvex ones in both directions rather than only in one. We state the resulting condition only for the special case of the welfare maximization problem, whose constraints are equalities, in \Cref{subsec: probabilistic market}, and do not pursue the general case here.
\end{remark}

Let us start by studying the value function of the convex subproblem where the variables $z_i$ of the nonconvex indices $i\in\mathcal{N}$ are fixed. 
Let us denote $\mathbf{\bar{z}}= (\bar{z}_i)_{i\in\mathcal{N}}\in \prod_{i\in\mathcal{N}} V_i$:
\begin{subequations}\label{eq: convex value problem}
    \begin{align}
        v_{\mathbf{\bar{z}}}(b) \; := \; \min_{(z_i)_{i\in\mathcal{C}}} & \quad \sum_{i\in\mathcal{N}} \bar{z}_i^0 + \sum_{i\in\mathcal{C}} z_i^0 \\
        \text{s.t.} & \quad \sum_{i\in\mathcal{N}} \bar{z}_i^k + \sum_{i\in\mathcal{C}} z_i^k \le b^k \quad \forall k\in\mathcal{K} \\
        & \quad z_i \in V_i \quad \forall i\in\mathcal{C} .
    \end{align}
\end{subequations}
    
We know that this value function is Lipschitz continuous under Slater's condition.

\begin{definition}[Slater's Condition]\label{def: slaters condition}
    For a given $b\in\mathbb{R}^m$ and $\mathbf{\bar{z}}\in\mathbb{R}^{\card{\mathcal{N}}(m+1)}$, we say that $v_{\mathbf{\bar{z}}}(b)$ satisfies Slater's condition if there exists a $(z_i)_{i\in\mathcal{C}}\in \prod_{i\in\mathcal{C}} V_i$ with $$\sum_{i\in\mathcal{N}} \bar{z}_i^k + \sum_{i\in\mathcal{C}} z_i^k \; < \; b^k \quad \forall k\in\mathcal{K}.$$
\end{definition}  

\begin{lemma}\label{lemma: lipschitz convex function}
If $v_{\mathbf{\bar{z}}}(b)$ satisfies Slater's condition for a given $\mathbf{\bar{z}}\in\mathbb{R}^{\card{\mathcal{N}}(m+1)}$ and all $b$ in a compact set $\mathcal{B}\subset\mathbb{R}^m$ then $v_{\mathbf{\bar{z}}}(b)$ is Lipschitz continuous over $b\in\mathcal{B}$ with some constant $L$.
\end{lemma}
\begin{proof}
Because~\eqref{eq: convex value problem} is a convex optimization problem, $v_{\mathbf{\bar{z}}}(b)$ is a convex function.
Every convex function is locally Lipschitz continuous on the relative interior of its domain.
By Slater's condition the set $\mathcal{B}$ lies in the relative interior of the domain of $v_{\mathbf{\bar{z}}}(b)$ and hence $v_{\mathbf{\bar{z}}}(b)$ is locally Lipschitz on $\mathcal{B}$. Because $\mathcal{B}$ is compact, $v_{\mathbf{\bar{z}}}(b)$ is (globally) Lipschitz on $\mathcal{B}$.
\end{proof}

The value function $v(b)$ of the whole problem and the value function $v_{\mathbf{\bar{z}}}(b)$ of the convex subproblem are related as follows:
\begin{equation*}
    v(b) \; = \; \min_{\mathbf{\bar{z}}} \; v_{\mathbf{\bar{z}}}(b) \;\; \text{s.t.} \;\; \mathbf{\bar{z}} \in \prod_{i\in\mathcal{N}} V_i .
\end{equation*}
Since $V_i=\im{f_i}+\mathbb{R}^{m+1}_{\ge0}$ (\Cref{subsec: preliminaries}), every $\mathbf{\bar{z}}\in\prod_{i\in\mathcal{N}}V_i$ can be written as $\mathbf{\bar{z}}=\mathbf{w}+\mathbf{d}$ with $\mathbf{w}\in\prod_{i\in\mathcal{N}}\im{f_i}$ and $\mathbf{d}=(d_i)_{i\in\mathcal{N}}\ge0$. Replacing $\mathbf{\bar{z}}$ by $\mathbf{w}$ lowers the objective of~\eqref{eq: convex value problem} by $\sum_{i\in\mathcal{N}}d_i^0$ and relaxes each constraint by $\sum_{i\in\mathcal{N}}d_i^k$, so that $v_{\mathbf{w}}(b)\le v_{\mathbf{\bar{z}}}(b)$. As $\im{f_i}\subseteq V_i$, we may therefore restrict the minimization to the images, and the minimum over $\prod_{i\in\mathcal{N}}\im{f_i}$ is attained whenever the one over $\prod_{i\in\mathcal{N}}V_i$ is: 
\begin{equation}\label{eq: nested value function}
    v(b) \; = \; \min_{\mathbf{\bar{z}}} \; v_{\mathbf{\bar{z}}}(b) \;\; \text{s.t.} \;\; \mathbf{\bar{z}} \in \prod_{i\in\mathcal{N}} \im{f_i} .
\end{equation}

In \Cref{lemma: lipschitz convex function}, we assumed a point $\mathbf{\bar{z}}$ was given to us. 
If we assume the condition in \Cref{def: slaters condition} holds for all points $\mathbf{\bar{z}}\in\prod_{i\in\mathcal{N}} \cl{\im{f_i}}$, then we can even say that the value function $v(b)$ of the original problem is Lipschitz.

\begin{lemma}\label{lemma: lipschitz value function}
If $v_{\mathbf{\bar{z}}}(b)$ satisfies Slater's condition for all $\mathbf{\bar{z}}\in\prod_{i\in\mathcal{N}} \cl{\im{f_i}}$ and all $b$ in a compact set $\mathcal{B}\subset\mathbb{R}^m$ then $v(b)$ is Lipschitz continuous over $b\in\mathcal{B}$ with some constant $L$.
\end{lemma}
\begin{proof}
Let us denote $\mathcal{Z}=\prod_{i\in\mathcal{N}} \im{f_i}$.
We first show that the functions $v_{\mathbf{\bar{z}}}$, $\mathbf{\bar{z}}\in\mathcal{Z}$, are Lipschitz over $\mathcal{B}$ with a common constant $L$. Let $\phi(c)$ denote the optimal value of~\eqref{eq: convex value problem} without nonconvex indices, i.e., with $\mathcal{N}=\emptyset$ and right-hand side $c\in\mathbb{R}^m$. Then
\begin{equation*}
    v_{\mathbf{\bar{z}}}(b) \; = \; \sum_{i\in\mathcal{N}} \bar{z}_i^0 \; + \; \phi\Big(\big(b^k-\textstyle\sum_{i\in\mathcal{N}}\bar{z}_i^k\big)_{k\in\mathcal{K}}\Big) ,
\end{equation*}
so all functions $v_{\mathbf{\bar{z}}}$ are translates of the single convex function $\phi$. As $b$ ranges over the compact set $\mathcal{B}$ and $\mathbf{\bar{z}}$ over the compact set $\cl{\mathcal{Z}}\subseteq[-\gamma,\gamma]^{\card{\mathcal{N}}(m+1)}$, the argument of $\phi$ ranges over a compact set, at every point of which $\phi$ satisfies Slater's condition by assumption. By \Cref{lemma: lipschitz convex function} (with $\mathcal{N}=\emptyset$), $\phi$ is Lipschitz on this set with some constant $L$, and hence $\lvert v_{\mathbf{\bar{z}}}(b_1)-v_{\mathbf{\bar{z}}}(b_2)\rvert\le L\norm{b_1-b_2}_\infty$ for all $b_1,b_2\in\mathcal{B}$ and $\mathbf{\bar{z}}\in\mathcal{Z}$.

Now consider two points $b_1,b_2\in\mathcal{B}$.
Because $v_{\mathbf{\bar{z}}}(b)$ satisfies Slater's condition for all $\mathbf{\bar{z}}\in\mathcal{Z}$ and all $b\in\mathcal{B}$, we know that $v_{\mathbf{\bar{z}}}(b)<\infty$ for all $\mathbf{\bar{z}}\in\mathcal{Z}$ and $b\in\mathcal{B}$. By \Cref{assumption: random functions}(a) and the argument preceding~\eqref{eq: nested value function}, the minimum in~\eqref{eq: nested value function} is attained for every $b\in\mathcal{B}$.
Hence, let us write
\begin{equation*}
    \mathbf{\bar{z}}(b_2) \in  \argmin_{\mathbf{\bar{z}} \in \mathcal{Z}} \; v_{\mathbf{\bar{z}}}(b_2) .
\end{equation*}
Since $v_{\mathbf{\bar{z}}(b_2)}(b_1)<\infty$, the point $\mathbf{\bar{z}}(b_2)$ gives an upper bound for $v(b_1)$ by the definition of the minimum: 
$$v(b_1) = \min_{\mathbf{\bar{z}} \in \mathcal{Z}} v_{\mathbf{\bar{z}}}(b_1) \le v_{\mathbf{\bar{z}}(b_2)}(b_1).$$

Consequently, we can write:
\begin{align*}
    v(b_1) - v(b_2) & \ = \ \min_{\mathbf{\bar{z}} \in \mathcal{Z}} v_{\mathbf{\bar{z}}}(b_1) - \min_{\mathbf{\bar{z}} \in \mathcal{Z}} v_{\mathbf{\bar{z}}}(b_2) \\
    & \ \le \ v_{\mathbf{\bar{z}}(b_2)}(b_1) -  v_{\mathbf{\bar{z}}(b_2)}(b_2)  \\
    & \ \le \ L \cdot \norm{b_1-b_2}_\infty ,
\end{align*}
where the last line follows from the common Lipschitz bound above. Exchanging the roles of $b_1$ and $b_2$ gives $v(b_2)-v(b_1)\le L\norm{b_1-b_2}_\infty$, which proves the statement.
\end{proof}

The reason why the value function $v(b)$ might not be Lipschitz is that with an incremental change $\epsilon$ in $b$, a point $\mathbf{\bar{z}}$ might suddenly become feasible at $b+\epsilon$, which was not feasible at $b$.
If we assume that all points $\mathbf{\bar{z}}$ are feasible for all $b$, this sudden jump no longer occurs, ensuring continuity. The following example illustrates this.

\begin{example}\label{example: discontinuous value function}
    Consider the optimization problem
    \begin{align*}
        \max & \quad x_1  \\
        \text{s.t.} & \quad x_1 + x_2 \le b \\
        & \quad x_1 \in [0,1] \cup [3,4] \\
        & \quad x_2 \in[0,4].
    \end{align*}
    The assumption in \Cref{lemma: lipschitz value function} is not fulfilled at $b<3$ as $x_1\ge3$ is not feasible for any $x_2\in[0,4]$, leading to a discontinuous value function at the point $b=3$ when the point $x_1=3$ suddenly becomes feasible. 
    If we have $x_2\in[-4,4]$ instead, all $x_1 \in [0,1] \cup [3,4]$ are feasible for $b>0$ ensuring a continuous value function. \Cref{fig:decision_space_analysis} illustrates this example.
\end{example}

A similar argument as in \Cref{lemma: lipschitz value function} is formalized in the concept of \textit{complete continuous recourse} for stochastic mixed-integer problems introduced in~\citet{zou2019stochastic} and discussed extensively in \citet{ahmed2022stochastic}. Our argument in \Cref{lemma: lipschitz value function} can be seen as a generalization of that concept.

\begin{figure}[t]
\centering
\begin{tikzpicture}
    % =========================================================================
    \begin{scope}[shift={(0,0)}]
        \begin{axis}[
            width=7cm, height=7cm,
            xmin=-0.5, xmax=4.5,
            ymin=-0.5, ymax=4.5,
            xtick={0,1,2,3,4}, ytick={0,1,2,3,4},
            grid=both, grid style={cbMidnightBlue!10},
            axis lines=box, 
            tick label style={font=\small},
            xlabel={$x_1$},
            x label style={at={(axis description cs:0.5,-0.12)}, anchor=north, font=\small},
            ylabel={$x_2$},
            y label style={at={(axis description cs:-0.12,0.5)}, anchor=south, rotate=0, font=\small}
        ]
            % Draw the two disjoint domain strips for x_1
            \draw[dashed, cbMidnightBlue!40, thick] (axis cs:0,0) -- (axis cs:0,4) -- (axis cs:1,4) -- (axis cs:1,0) -- cycle;
            \draw[dashed, cbMidnightBlue!40, thick] (axis cs:3,0) -- (axis cs:3,4) -- (axis cs:4,4) -- (axis cs:4,0) -- cycle;
            % Shaded feasible region for b = 2
            \addplot[fill=cbDarkNavy!20, draw=none] coordinates {
                (0,0) (0,2) (1,1) (1,0)
            };
            % Constraint line for b = 2 (x_1 + x_2 = 2)
            \addplot[draw=cbDarkNavy, ultra thick, domain=-0.5:4.5] {2 - x};
            % Optimal point (1,0) for b = 2
            \addplot[only marks, mark=*, mark size=3pt, mark options={fill=cbDeepSteel, draw=cbMidnightBlue}] coordinates {(1,0)};
        \end{axis}
    \end{scope}
    % =========================================================================
    \begin{scope}[shift={(7.2,0)}]
        \begin{axis}[
            width=7cm, height=7cm,
            xmin=-0.5, xmax=4.5,
            ymin=-0.5, ymax=4.5,
            xtick={0,1,2,3,4}, ytick={0,1,2,3,4},
            grid=both, grid style={cbMidnightBlue!10},
            axis lines=box, 
            tick label style={font=\small},
            xlabel={$x_1$},
            x label style={at={(axis description cs:0.5,-0.12)}, anchor=north, font=\small},
            ylabel={$x_2$},
            y label style={at={(axis description cs:-0.12,0.5)}, anchor=south, rotate=0, font=\small}
        ]
            % Draw the two disjoint domain strips for x_1
            \draw[dashed, cbMidnightBlue!40, thick] (axis cs:0,0) -- (axis cs:0,4) -- (axis cs:1,4) -- (axis cs:1,0) -- cycle;
            \draw[dashed, cbMidnightBlue!40, thick] (axis cs:3,0) -- (axis cs:3,4) -- (axis cs:4,4) -- (axis cs:4,0) -- cycle;
            % Shaded feasible region for b = 3 (strip x_1 in [0,1])
            \addplot[fill=cbDarkNavy!20, draw=none] coordinates {
                (0,0) (0,3) (1,2) (1,0)
            };
            % Constraint line for b = 3 (x_1 + x_2 = 3)
            \addplot[draw=cbDarkNavy, ultra thick, domain=-0.5:4.5] {3 - x};
            % Highlight the sudden activation of point (3,0)
            \addplot[only marks, mark=*, mark size=3pt, mark options={fill=cbDeepSteel, draw=cbMidnightBlue}] coordinates {(3,0)};
        \end{axis}
    \end{scope}
\end{tikzpicture}
\caption{Illustration of \Cref{example: discontinuous value function} for $b=2$ (left) and $b=3$ (right). At $b=2$, the optimal value is attained at $x_1=1$; at $b=3$, the point $x_1=3$ becomes feasible and the optimal value jumps.}\label{fig:decision_space_analysis}
\end{figure}

In \Cref{theorem: non-zero duality gap}, we have assumed that $v(b)$ is Lipschitz over the compact box $\mathcal{B}\subset\mathbb{R}_{\ge0}^m$ defined in equation~\eqref{eq: box}.
We can use \Cref{lemma: lipschitz value function} to obtain a simple sufficient condition for this to be true using the bounded image $\im{f_i}\subseteq[-\gamma,\gamma]^{m+1}$ in \Cref{assumption: random functions}(a).

\begin{lemma}\label{lemma: sufficient condition}
    $v(b)$ is Lipschitz continuous over $\mathcal{B} \; = \; \big\{ b\in\mathbb{R}_{\ge0}^m \ \big| \ \norm{b}_{\infty} \le (m+1) \cdot 2\gamma \big\}$ if 
    \begin{equation}\label{eq: condition}
        \sum_{i\in\mathcal{C}} \; \min_{x_i} \; \max_{k\in\mathcal{K}} \;  f_i^k(x_i) < - \card{\mathcal{N}} \gamma  .
    \end{equation} 
\end{lemma}
\begin{proof}
\Cref{lemma: lipschitz value function} tells us that $v(b)$ is Lipschitz over the compact box $\mathcal{B}\subset\mathbb{R}_{\ge0}^m$ if
\begin{equation*}
    \forall (\bar{z}_i)_{i\in\mathcal{N}}\in\prod_{i\in\mathcal{N}} \cl{\im{f_i}} \quad \exists (z_i)_{i\in\mathcal{C}}\in \prod_{i\in\mathcal{C}} V_i \quad \text{s.t.} \quad \sum_{i\in\mathcal{N}} \bar{z}_i^k + \sum_{i\in\mathcal{C}} z_i^k < 0 \quad \forall k\in\mathcal{K}
\end{equation*}
where the 0 on the right-hand side comes from $b\ge0$ for all $b\in\mathcal{B}\subset\mathbb{R}_{\ge0}^m$.
Because $\cl{\im{f_i}}\subseteq[-\gamma,\gamma]^{m+1}$, we know that a sufficient condition for this statement to be true is
\begin{equation*}
        \exists (z_i)_{i\in\mathcal{C}}\in \prod_{i\in\mathcal{C}} V_i \quad \text{s.t.} \quad \sum_{i\in\mathcal{C}} z_i^k < - \card{\mathcal{N}} \gamma \quad \forall k\in\mathcal{K} .
\end{equation*}    
We can replace the for-all statement by the max operator over $k\in\mathcal{K}$:
\begin{equation*}
        \exists (z_i)_{i\in\mathcal{C}}\in \prod_{i\in\mathcal{C}} V_i \quad \text{s.t.} \quad \max_{k\in\mathcal{K}} \sum_{i\in\mathcal{C}} z_i^k < - \card{\mathcal{N}} \gamma .
\end{equation*}    
Next, we can replace the exists statement $\exists$ by the minimum:
\begin{equation*}
        \min_{(z_i)_{i\in\mathcal{C}}\in \prod_{i\in\mathcal{C}} V_i} \; \max_{k\in\mathcal{K}} \; \sum_{i\in\mathcal{C}} z_i^k < - \card{\mathcal{N}} \gamma .
\end{equation*} 
Finally, we can replace the epigraph-projection $V_i$ by the function $f_i$:
\begin{equation*}
        \min_{(x_i)_{i\in\mathcal{C}}} \; \max_{k\in\mathcal{K}} \; \sum_{i\in\mathcal{C}} f_i^k(x_i) < - \card{\mathcal{N}} \gamma  .
\end{equation*}    
Since $\max_{k\in\mathcal{K}}\sum_{i\in\mathcal{C}} f_i^k(x_i)\le\sum_{i\in\mathcal{C}}\max_{k\in\mathcal{K}} f_i^k(x_i)$ for every $(x_i)_{i\in\mathcal{C}}$, and the minimum of a separable sum equals the sum of the minima, the left-hand side is bounded above by $\sum_{i\in\mathcal{C}}\min_{x_i}\max_{k\in\mathcal{K}} f_i^k(x_i)$, and the statement follows.
\end{proof}

Condition~\eqref{eq: condition} tells us that if the i.i.d. functions $f_i,i\in\mathcal{C}$ can become sufficiently negative in all their $m$-components, then we can absorb the nonconvexity of functions $f_i,i\in\mathcal{N}$.
To understand when this is the case, we need to understand the quantity $\min_{x_i} \max_{k\in\mathcal{K}}  f_i^k(x_i)$ better, which tells us how negative $f_i$ can become over its $m$-component functions. 
For this, let us denote
\begin{equation*}
    \psi_i := \; \min_{x_i} \max_{k\in\mathcal{K}}  f_i^k(x_i) .
\end{equation*}
By \Cref{assumption: random functions}(b) and~(d), the functions $f_i,i\in\mathcal{C}$, are i.i.d., and hence so are the random variables $\psi_i,i\in\mathcal{C}$:

\begin{lemma}\label{lemma: iid psi}
    The random variables $\psi_i$, $i\in\mathcal{C}$, are i.i.d.
\end{lemma}
\begin{proof}
    For $i\in\mathcal{C}$ we have $\co{\epi{f_i}}=\epi{f_i}$ by \Cref{assumption: random functions}(b), so the epigraphs $\epi{f_i}$, $i\in\mathcal{C}$, are i.i.d.\ by \Cref{assumption: random functions}(d). Moreover, $\psi_i=\Psi(\epi{f_i})$ for the deterministic map
    \begin{equation*}
        \Psi(F) \; := \; \min \big\{ \max_{k\in\mathcal{K}} z^k \; \big| \;
        (x,z)\in F \big\} ,
    \end{equation*}
    which is measurable, and whose minimum is attained by compactness of $\dom{f_i}$ and lower semicontinuity of $f_i$. Applying the same map to i.i.d.\ random closed sets preserves independence and equality of laws.
\end{proof}

By \Cref{assumption: random functions}(a), all $\psi_i$ lie in $[-\gamma,\gamma]$.
We can now characterize the likelihood of $v(b)$ being Lipschitz over $\mathcal{B}$ solely through the random variables $\psi_i$, $i\in\mathcal{C}$.
A simple but informative case is the following.

\begin{proposition}\label{proposition: simple characterisation lipschitz}
    If $\psi_i<-\frac{\card{\mathcal{N}}\gamma}{\card{\mathcal{C}}}$ for all $i\in\mathcal{C}$ with probability 1 then $v(b)$ is Lipschitz over $\mathcal{B}$ with probability~1.
\end{proposition}
\begin{proof}
    Follows directly from \Cref{lemma: sufficient condition}.
\end{proof}

\Cref{proposition: simple characterisation lipschitz} tells us that if for every realization, the functions $f_i$ can take some negative value smaller than $-\frac{\card{\mathcal{N}}\gamma}{\card{\mathcal{C}}}$ on all $m$ coordinates then $v(b)$ is Lipschitz.
The constant $\frac{\card{\mathcal{N}}\gamma}{\card{\mathcal{C}}}$ decreases with the number of convex functions and increases with the number of nonconvex ones. 

To characterize the probability of $v(b)$ being Lipschitz beyond this simple case, we need to understand the distribution of the sum $\sum_{i\in\mathcal{C}} \psi_i$ on its lower tail $\sum_{i\in\mathcal{C}} \psi_i < - \card{\mathcal{N}} \gamma$. 
For this, we can use the fact that the sum of i.i.d. random variables approximately follows a normal distribution.
The Berry--Esseen theorem formalizes this~\citep{berry1941accuracy,esseen1942liapounoff}. For our purposes of lower-bounding the tail probability, it can be stated as follows:

\begin{lemma}[Berry–Esseen theorem]\label{lemma: berry-esseen}
Let $X_1, \dots, X_n$ be i.i.d. random variables with $\var{X_i}>0$ and finite third moments $\rho=\mathbb{E}[|X_i - \E{X_i}|^3] < \infty$. Then 
\begin{equation}
    \Pr\left( \sum_{i=1}^n X_i - \mathbb{E}[\sum_{i=1}^n X_i] \le t \right) \ge \Phi\left( \frac{t}{\sqrt{\var{X_i}} \sqrt{n}} \right) - \frac{G \rho}{\sqrt{\var{X_i}}^3 \sqrt{n}}
\end{equation}
where $\Phi(x)$ is the cumulative distribution function (CDF) of the standard normal distribution with expected value 0 and variance 1, and $G$ is a constant bounded by $G < 0.4748$~\citep{shevtsova2011absolute}.
\end{lemma}

\begin{figure}[tb]
\centering
\begin{tikzpicture}[scale=0.8]
    \begin{axis}[
        width=11cm, height=11cm,
        xmin=-2.3, xmax=2.3,
        ymin=-0.05, ymax=1.05,
        xtick={-3, -2, -1, 0, 1, 2, 3},
        ytick={0.2, 0.6, 1.0},
        % Replicating the center-intersecting axis style from the image
        axis lines=middle,
        inner axis line style={-},
        axis line style={thick},
        tick label style={font=\small},
        % Removing background grids to replicate image simplicity
        grid=none,
        % Legend positioning and styling inside the clean plot
        legend style={
            at={(axis description cs:0,0.95)}, 
            anchor=north west, 
            font=\small, 
            draw=cbMidnightBlue!20, 
            text=cbMidnightBlue
        }
    ]
        % 1. Approximating Normal CDF \Phi(x) - Solid Blue
        \addplot[
            draw=cbDarkNavy, 
            ultra thick, 
            domain=-3.3:3.3, 
            samples=100
        ] {1/(1+exp(-1.2*x)) + 0.02}; 
        \addlegendentry{Normal CDF $\Phi$};
        % 2. True CDF F_n(x) - Black and Dashed with pronounced gap deviation
        \addplot[
            draw=black, 
            dashed,
            ultra thick, 
            smooth,
            domain=-3.3:3.3,
            samples=100
        ] {1/(1+exp(-1.6*x)) + 0.10*x*exp(-0.35*x*x) + 0.02};
        \addlegendentry{CDF of $\sum_{i=1}^n X_i$};
        % Double-headed arrow at the lower tail (around x = -1.5)
        \draw[{Stealth[scale=1]}-{Stealth[scale=1]}, cbMidnightBlue, thick] 
            (axis cs:-1.5, 0.15) -- (axis cs:-1.5, 0.04);
        % Double-headed arrow at the upper tail (around x = 1.5)
        \draw[{Stealth[scale=1]}-{Stealth[scale=1]}, cbMidnightBlue, thick] 
            (axis cs:1.5, 0.995) -- (axis cs:1.5, 0.885);
    \end{axis}
\end{tikzpicture}
\caption{Illustration of the Berry–Esseen theorem.}\label{fig:berry_esseen_with_legend}
\end{figure}

The Berry–Esseen theorem tells us that the CDF of $\sum_{i\in\mathcal{C}} \psi_i$ is larger than the CDF of the normal distribution with the same mean and variance minus some constant $\frac{G \rho}{\sqrt{\var{X_i}}^3 \sqrt{n}}$, which goes to zero with increasing number of random variables. 
Evaluating the CDF at $-\card{\mathcal{N}} \gamma$ we obtain:

\begin{theorem}\label{theorem: probability lipschitz}
Let $\var{\psi_i}>0$ for $i\in\mathcal{C}$. Then
\begin{equation*}
    \Pr\big[ \textup{$v(b)$ is Lipschitz over $\mathcal{B}$} \big] \ge \Phi\left( \frac{-\card{\mathcal{N}}\gamma-\card{\mathcal{C}} \E{\psi_i}}{\sqrt{\var{\psi_i}} \sqrt{\card{\mathcal{C}}}} \right) - \frac{G \rho}{\sqrt{\var{\psi_i}}^3 \sqrt{\card{\mathcal{C}}}} ,
\end{equation*}
where $\rho:=\E{|\psi_i - \E{\psi_i}|^3}\le(2\gamma)^3<\infty$. %By \Cref{lemma: iid psi}, the quantities $\E{\psi_i}$, $\var{\psi_i}$, and $\rho$ do not depend on $i\in\mathcal{C}$.
\end{theorem}
\begin{proof}
By \Cref{lemma: iid psi}, the random variables $\psi_i$, $i\in\mathcal{C}$, are i.i.d., and $\rho<\infty$ holds because $\psi_i\in[-\gamma,\gamma]$. Applying \Cref{lemma: berry-esseen} to $\sum_{i\in\mathcal{C}} \psi_i$ therefore gives
\begin{equation*}
    \Pr\left( \sum_{i\in\mathcal{C}} \psi_i \le t + \card{\mathcal{C}} \E{\psi_i} \right) \ge \Phi\left( \frac{t}{\sqrt{\var{\psi_i}} \sqrt{\card{\mathcal{C}}}} \right) - \frac{G \rho}{\sqrt{\var{\psi_i}}^3 \sqrt{\card{\mathcal{C}}}} .
\end{equation*}
\Cref{lemma: sufficient condition} requires the strict inequality $\sum_{i\in\mathcal{C}}\psi_i<-\card{\mathcal{N}}\gamma$. We therefore evaluate at $t=-\card{\mathcal{N}}\gamma-\card{\mathcal{C}} \E{\psi_i}-\epsilon$ with $\epsilon>0$ and let $\epsilon\downarrow0$: the events $\{\sum_{i\in\mathcal{C}}\psi_i\le-\card{\mathcal{N}}\gamma-\epsilon\}$ increase to $\{\sum_{i\in\mathcal{C}}\psi_i<-\card{\mathcal{N}}\gamma\}$, so their probabilities converge by continuity of $\Pr$ from below, while the right-hand side converges by continuity of $\Phi$. This gives
\begin{equation*}
    \Pr\left( \sum_{i\in\mathcal{C}} \psi_i < -\card{\mathcal{N}}\gamma \right) \ge \Phi\left( \frac{-\card{\mathcal{N}}\gamma-\card{\mathcal{C}} \E{\psi_i}}{\sqrt{\var{\psi_i}} \sqrt{\card{\mathcal{C}}}} \right) - \frac{G \rho}{\sqrt{\var{\psi_i}}^3 \sqrt{\card{\mathcal{C}}}} .
\end{equation*}
The statement then follows from \Cref{lemma: sufficient condition}.
\end{proof}

From \Cref{theorem: probability lipschitz}, we obtain the asymptotic case in which the number of convex functions grows while the number of nonconvex ones stays fixed.

\begin{corollary}\label{corollary: asymptotic lipschitz}
Let $\var{\psi_i}>0$ and $\E{\psi_i}<0$ for $i\in\mathcal{C}$, and let $\card{\mathcal{N}}$ be fixed. Then
\begin{equation*}
    \lim_{\card{\mathcal{C}}\to\infty} \; \Pr\big[ \textup{$v(b)$ is Lipschitz over $\mathcal{B}$} \big] \; = \; 1 .
\end{equation*}
\end{corollary}
\begin{proof}
Since $\E{\psi_i}<0$, the numerator of the argument of $\Phi$ in \Cref{theorem: probability lipschitz} equals $\card{\mathcal{C}}\,\lvert\E{\psi_i}\rvert-\card{\mathcal{N}}\gamma$, so the argument reads
\begin{equation*}
    \frac{\card{\mathcal{C}}\,\lvert\E{\psi_i}\rvert-\card{\mathcal{N}}\gamma}{\sqrt{\var{\psi_i}} \sqrt{\card{\mathcal{C}}}} ,
\end{equation*}
which tends to $+\infty$ as $\card{\mathcal{C}}\to\infty$ for fixed $\card{\mathcal{N}}$, since the numerator grows linearly in $\card{\mathcal{C}}$ while the denominator grows only as $\sqrt{\card{\mathcal{C}}}$. Hence $\Phi(\cdot)\to1$, and the subtracted term tends to $0$ because $\rho$ and $\var{\psi_i}$ do not depend on $\card{\mathcal{C}}$.
\end{proof}

\subsection{Summary}\label{subsec: summary}

In summary, \Cref{theorem: zero duality gap,theorem: non-zero duality gap,theorem: probability lipschitz} together characterize the distribution of the duality gap in two steps. The bound on the probability of a zero duality gap in \Cref{theorem: zero duality gap} holds unconditionally. The bound on the entire distribution in \Cref{theorem: non-zero duality gap}, in contrast, requires a Lipschitz-continuous value function, and \Cref{theorem: probability lipschitz} bounds from below the probability that this requirement is met. Both bounds converge to 1 as the number of convex functions $\card{\mathcal{C}}$ grows, so in a nearly convex problem, we can expect the bound of \Cref{theorem: non-zero duality gap} to be applicable and the duality gap to be small.

\section{Walrasian Equilibrium}\label{sec: equilibrium}

As mentioned in the introduction, the duality gap is of particular economic significance as it characterizes the existence of a Walrasian equilibrium: a market has an equilibrium if and only if the duality gap of the corresponding welfare maximization problem vanishes. The results of \Cref{sec: duality gap} therefore apply directly to this special case, which we study in more detail below.

Let us consider a market with $\mathcal{I}=\{1,\ldots,I\}$ agents and $\mathcal{K}=\{1,\ldots,m\}$ commodities. 
Let us denote by $x_{i}=(x_{ik})_{k\in\mathcal{K}}$ a vector of commodities consumed (if $x_{ik}>0$) or supplied (if $x_{ik}<0$) by agent $i$.
The function $u_i:\mathbb{R}^{m}\to\mathbb{R}\cup\{-\infty\}$ is the utility function of agent $i$ and $u_i(x_i)$ denotes the utility agent $i$ gains from consuming/supplying bundle $x_i$. The domain $\dom{u_i} := \big\{ x_i\in\mathbb{R}^m \mid u_i(x_i) > -\infty \big\}$ contains all the bundles feasible to agent $i$.

A Walrasian equilibrium in this market is an allocation $x^\ast=(x^\ast_1,\ldots,x^\ast_I)$ and a price vector $\lambda^\ast=(\lambda_1^\ast,\ldots,\lambda_m^\ast)$ so that every agent is maximizing their utility $u_i(x_i^\ast)$ minus payments $\lambda^\ast x_i^\ast$ with trade $x_i^\ast$ and supply and demand of commodities is balanced $\sum_{i\in\mathcal{I}} x^\ast_{i} = 0$.

\begin{definition}[Walrasian Equilibrium] \label{def: walrasian equilibrium}
The tuple $(x^\ast, \lambda^\ast)$ is a Walrasian equilibrium if $\sum_{i\in\mathcal{I}} x^\ast_{i} = 0$ and $x_i^\ast \in \argmax \; u_i(x_i) - \lambda^\ast x_i$ for all $i\in\mathcal{I}$. 
\end{definition}

In what follows, we discuss the existence of a Walrasian equilibrium by studying the market's welfare maximization problem and its duality gap.
Throughout, the utility functions $u_i$ are random functions on $(\Omega,\mathcal{F},\Pr)$, and we make assumptions analogous to \Cref{assumption: random functions}.

\needspace{6\baselineskip}
\begin{assumption}\label{assumption: market}
The following holds for almost every $\omega\in\Omega$ and all $i\in\mathcal{I}$.
\begin{enumerate}[label=(\alph*)]
    \item \emph{Regular}: The domain $\dom{u_i}$ is nonempty and compact with $\dom{u_i}\subseteq[-\beta,\beta]^{m}$, $u_i$ is upper semicontinuous, has bounded image $\im{u_i}\subseteq[-\gamma,\gamma]$, and there exists a feasible point $\bar{x} \in \prod_{i\in\mathcal{I}} \relinterior{\co{\dom{u_i}}}$ with $\sum_{i\in\mathcal{I}} \bar{x}_i = 0$.
    \item \emph{Partially nonconvex}: For indices $i\in\mathcal{C}$, the functions $u_i$ are concave.
    \item \emph{Random}: The map $\omega\mapsto\hyp{u_i}$ is measurable, i.e., a random closed set, and for all indices $i\in\mathcal{I}$, the concave envelopes $\cav{u_i}$ are i.i.d.
\end{enumerate}
\end{assumption}

Parts~(b) and~(c) are used only in \Cref{subsec: probabilistic market}. In contrast to \Cref{assumption: random functions}(a), part~(a) requires the feasible point to lie in a relative interior, namely in that of the convex hull $\co{\dom{u_i}}$, which is the domain of the concave envelope $\cav{u_i}$ (\Cref{subsec: convexified market}). By Slater's condition applied to the convexified market~\eqref{eq: convex welfare maximization}, this guarantees that the dual optimum in~\eqref{eq: dual welfare maximization} is attained, i.e., that a price vector $\lambda^\ast$ exists. Such a price is essential for the economic analysis, whereas the analysis in \Cref{sec: duality gap} required only a solution to the geometric dual~\eqref{eq: dual epigraph} and did not require the existence of optimal dual multipliers. Part~(c) plays the role of \Cref{assumption: random functions}(c) and~(d): since $\co{\hyp{u_i}}=\hyp{\cav{u_i}}$ (\Cref{subsec: convexified market}), the concave envelopes are random closed sets, and the events considered in \Cref{subsec: probabilistic market} are measurable, as in \Cref{subsec: assumptions}.

\subsection{Welfare Maximization Problem}\label{subsec: welfare max problem}

The welfare maximization problem asks how to allocate bundles of commodities to agents so as to maximize social welfare which is the sum of individual utilities. This optimization problem can be stated as follows:
\begin{subequations}\label{eq: welfare maximization}
    \begin{align}
        v^\ast \; := \; \max_x & \quad \sum_{i\in\mathcal{I}} u_i(x_i) \\
        \text{s.t.} & \quad \sum_{i\in\mathcal{I}} x_i = 0 .
    \end{align}
\end{subequations}
The constraints ensure that the amount of commodities supplied and consumed in the market is equal. 
The dual of this problem is given by:
\begin{align}\label{eq: dual welfare maximization}
   v^{\ast\ast} \; := \; \min_{\lambda} \; \max_x  \; \sum_{i\in\mathcal{I}} \big( u_i(x_i) - \lambda x_i \big) ,
\end{align}
where $\lambda\in\mathbb{R}^m$ is the Lagrange multiplier, which has the economic meaning of a commodity price vector.
A Walrasian equilibrium exists if and only if the welfare maximization problem in this market has a zero duality gap~\citep{mas1995microeconomic,bichler2021walrasian}. The proof is included here for completeness. 

\begin{proposition}\label{proposition: equilibrium existence duality gap}
        $(x^\ast, \lambda^\ast)$ is a Walrasian equilibrium if and only if $v^\ast=v^{\ast\ast}$, $x^\ast$ solves~\eqref{eq: welfare maximization}, and $\lambda^\ast$ solves~\eqref{eq: dual welfare maximization}.
\end{proposition}
\begin{proof}
Let us denote $g(\lambda) := \max_x  \; \sum_{i\in\mathcal{I}} \big( u_i(x_i) - \lambda x_i \big)$.

($\Leftarrow$) Suppose $v^\ast=v^{\ast\ast}$, $x^\ast$ solves~\eqref{eq: welfare maximization}, and $\lambda^\ast$ solves~\eqref{eq: dual welfare maximization}. 
Then 
$$\sum_{i\in\mathcal{I}} u_i(x_i^\ast) = v^\ast = v^{\ast\ast} = g(\lambda^\ast).$$
Since $x^\ast$ is feasible, $\sum_{i\in\mathcal{I}} x_i^\ast=0$, so $\sum_{i\in\mathcal{I}}\lambda^\ast x_i^\ast=0$ and therefore $$\sum_{i\in\mathcal{I}} u_i(x_i^\ast) = \sum_{i\in\mathcal{I}} \big[u_i(x_i^\ast)-\lambda^\ast x_i^\ast\big].$$
Combining the two equations gives:
\begin{align*}
    \sum_{i\in\mathcal{I}} \big[u_i(x_i^\ast)-\lambda^\ast x_i^\ast\big] = \sum_{i\in\mathcal{I}} \max_{x_i} \big[u_i(x_i) - \lambda^\ast x_i\big] .
\end{align*}
Each term on the left is bounded above by the corresponding term on the right, since $x_i^\ast$ is a feasible point of the $i$-th maximization. As the two sums are equal, each term must be equal individually: $u_i(x_i^\ast)-\lambda^\ast x_i^\ast = \max_{x_i} \big[u_i(x_i) - \lambda^\ast x_i\big]$ for every $i\in\mathcal{I}$, i.e., $x_i^\ast \in \argmax_{x_i} \; u_i(x_i) - \lambda^\ast x_i$. Together with $\sum_{i\in\mathcal{I}} x_i^\ast=0$, this shows $(x^\ast,\lambda^\ast)$ is a Walrasian equilibrium.

($\Rightarrow$) Suppose $(x^\ast,\lambda^\ast)$ is a Walrasian equilibrium. Summing the agent-optimality condition over $i\in\mathcal{I}$,
\begin{align*}
    \sum_{i\in\mathcal{I}} \big[u_i(x_i^\ast) - \lambda^\ast x_i^\ast\big] = \sum_{i\in\mathcal{I}} \max_{x_i} \big[u_i(x_i) - \lambda^\ast x_i\big] = g(\lambda^\ast) .
\end{align*}
Market clearing gives $\sum_{i\in\mathcal{I}} x_i^\ast=0$, so the left-hand side equals $\sum_{i\in\mathcal{I}} u_i(x_i^\ast)$, and thus $\sum_{i\in\mathcal{I}} u_i(x_i^\ast) = g(\lambda^\ast)$.
Since $x^\ast$ is feasible for~\eqref{eq: welfare maximization}, $\sum_{i\in\mathcal{I}} u_i(x_i^\ast) \le v^\ast$. By weak duality, $v^\ast \le v^{\ast\ast} \le g(\lambda^\ast)$. Combining these,
\begin{align*}
    \sum_{i\in\mathcal{I}} u_i(x_i^\ast) \; \le \; v^\ast \; \le \; v^{\ast\ast} \; \le \; g(\lambda^\ast) \; = \; \sum_{i\in\mathcal{I}} u_i(x_i^\ast) ,
\end{align*}
so every inequality is in fact an equality. Hence $x^\ast$ attains $v^\ast$ (i.e., solves~\eqref{eq: welfare maximization}), and $\lambda^\ast$ attains $v^{\ast\ast}=\min_\lambda g(\lambda)$ (i.e., solves~\eqref{eq: dual welfare maximization}), with $v^\ast=v^{\ast\ast}$.
\end{proof}

If the duality gap is nonzero, an equilibrium does not exist.
However, the duality gap can be used to establish some notion of distance from a Walrasian equilibrium. 
For this, we need the concept of \textit{lost opportunity cost}, which was popularized in the electricity market literature~\citep{stevens2024some,ahunbay2025pricing} but also used in theoretical economics under the name ``rationing loss''~\citep{milgrom2026walrasian}. 

\begin{definition}[Lost Opportunity Cost]\label{def: lost opportunity cost}
Let $(x',\lambda')$ be an allocation-price pair.
We call 
\begin{equation}
    \Gamma(x',\lambda') \; := \; \sum_{i\in\mathcal{I}} \; \big[ \max_{x_i} \left( u_i(x_i) - \lambda' x_i\right) \; - \; \left(u_i(x_i') - \lambda' x_i'\right) \big]
\end{equation}
the lost opportunity cost of $(x',\lambda')$.
\end{definition}

The lost opportunity cost measures the difference between the maximum possible utility agents could have gained at price $\lambda'$ if they could freely choose their bundle $x_i$, and the utility they gained under allocation $x_i'$.
That the notion of lost opportunity cost serves well as a measure of distance is confirmed by the following.

\begin{proposition}\label{prop: zero locs iff equilibrium}
    A Walrasian equilibrium exists if and only if there is a tuple $(x',\lambda')$ with $\sum_{i\in\mathcal{I}} x_i' = 0$ and $\Gamma(x',\lambda')=0$.
\end{proposition}
\begin{proof}
    Follows directly from the definition of lost opportunity costs.
\end{proof}

The following result links the lost opportunity cost to the duality gap $v^{\ast\ast} - v^\ast$.

\begin{proposition}\label{proposition: locs and duality gap}
    For all tuples $(x',\lambda')$ with $\sum_{i\in\mathcal{I}} x_i' = 0$, it holds that $\Gamma(x',\lambda') \ge v^{\ast\ast} - v^\ast$. 
    If $x'$ is a solution to~\eqref{eq: welfare maximization} and $\lambda'$ a solution to~\eqref{eq: dual welfare maximization} then $\Gamma(x',\lambda') = v^{\ast\ast} - v^\ast$.
\end{proposition}
\begin{proof}
Fix $(x',\lambda')$ with $\sum_{i\in\mathcal{I}} x_i' = 0$. By definition,
\begin{equation*}
    \Gamma(x',\lambda') \; = \; \sum_{i\in\mathcal{I}} \max_{x_i} \big(u_i(x_i)-\lambda' x_i\big) \; - \; \sum_{i\in\mathcal{I}} \big(u_i(x_i') - \lambda' x_i'\big) .
\end{equation*}
Since $\sum_{i\in\mathcal{I}} x_i' = 0$, the second sum equals $\sum_{i\in\mathcal{I}} u_i(x_i')$, while the first sum is precisely the dual objective~\eqref{eq: dual welfare maximization} evaluated at $\lambda'$. Hence
\begin{equation*}
    \Gamma(x',\lambda') \; = \; \Big[\max_x \sum_{i\in\mathcal{I}} \big(u_i(x_i)-\lambda' x_i\big)\Big] \; - \; \sum_{i\in\mathcal{I}} u_i(x_i') .
\end{equation*}
The bracketed term is at least $v^{\ast\ast}$, since $v^{\ast\ast}$ is by definition the minimum of this expression over all $\lambda$. Moreover, since $x'$ is feasible for~\eqref{eq: welfare maximization}, $\sum_{i\in\mathcal{I}} u_i(x_i') \le v^\ast$. Combining these two bounds gives $\Gamma(x',\lambda') \ge v^{\ast\ast} - v^\ast$.

If $x'$ solves~\eqref{eq: welfare maximization} and $\lambda'$ solves~\eqref{eq: dual welfare maximization}, both inequalities hold with equality: the bracketed term equals $v^{\ast\ast}$ and $\sum_{i\in\mathcal{I}} u_i(x_i') = v^\ast$. Hence $\Gamma(x',\lambda') = v^{\ast\ast} - v^\ast$.
\end{proof}

\Cref{proposition: locs and duality gap} tells us that the duality gap $v^{\ast\ast} - v^\ast$ is the smallest lost opportunity cost that a tuple $(x',\lambda')$ with $\sum_{i\in\mathcal{I}} x_i' = 0$ can achieve, i.e., it measures how close any such tuple can come to a Walrasian equilibrium. This smallest value is attained by taking a solution to~\eqref{eq: welfare maximization} as allocation $x'$ and a solution to~\eqref{eq: dual welfare maximization} as price $\lambda'$, a method known as \textit{convex hull pricing}~\citep{stevens2024some}.

Together, \Cref{proposition: equilibrium existence duality gap,proposition: locs and duality gap} show that analyzing the duality gap of the welfare maximization problem is a natural way to understand the existence of a Walrasian equilibrium and---if it does not exist---how close we can get to a Walrasian equilibrium when ``closeness'' is measured by lost opportunity costs.
In the following, we will derive the economic equivalents of \Cref{theorem: zero duality gap} and \Cref{theorem: non-zero duality gap} for the special case of the welfare maximization problem.
We will derive the results in a similar way, but replace the geometric dual~\eqref{eq: dual epigraph} and the set $V_i$ by the economically more meaningful concepts of the convexified market and demand sets. 

\subsection{Convexified Market}\label{subsec: convexified market}

We would like to obtain the economic equivalent to~\Cref{lemma: extremepoints}, which was the central geometric argument of our analysis in \Cref{sec: duality gap}.
To do so, we introduce a geometric version of the dual~\eqref{eq: dual welfare maximization}, analogous to the problem~\eqref{eq: dual epigraph} used throughout \Cref{sec: duality gap}.
The economic meaning of this geometric dual is the ``convexified market''~\citep{starr1969quasi,milgrom2026walrasian}.

Since we are dealing with a maximization problem instead of a minimization problem as in \Cref{sec: duality gap}, we will use the hypograph instead of the epigraph of a function. The hypograph of $u_i$ is given by
\begin{equation*}
    \hyp{u_i} = \left\{ (x_i,t) \; | \; x_i \in \mathbb{R}^m, t\le u_i(x_i) \right\} .
\end{equation*}
Because $u_i$ is a scalar-valued function in comparison to the vector-valued $f_i$ discussed in \Cref{sec: duality gap}, we can use the concept of the concave envelope, defined as the smallest concave function that overestimates $u_i$. That is, 
\begin{equation*}
    \cav{u_i}(x_i) = \sup \left \{ t \; | \; (x_i,t) \in \co{\hyp{u_i}} \right\} .
\end{equation*}
For the special case of a scalar-valued function, the convex hull of the hypograph is the same as the hypograph of the concave envelope. This does not necessarily hold for vector-valued functions \citep[cf.][Appendix~A]{lemarechal2001geometric}. 

\begin{lemma}
    The convex hull $\co{\hyp{u_i}}$ of the hypograph $\hyp{u_i}$ equals the hypograph $\hyp{\cav{u_i}}$ of the concave envelope $\cav{u_i}$.
\end{lemma}
\begin{proof}
    See \citet[Proposition~A.5]{lemarechal2001geometric}.
\end{proof}

Now we can define the convexified market as the same market as defined earlier, but with utility functions $\cav{u_i}$ instead of $u_i$.
The ``geometric'' equivalent to the dual~\eqref{eq: dual welfare maximization} is the welfare maximization problem in the convexified market:
\begin{subequations}\label{eq: convex welfare maximization}
    \begin{align}
        v^{\ast\ast} \; = \; \max_x & \quad \sum_{i\in\mathcal{I}} \cav{u_i}(x_i) \\
        \text{s.t.} & \quad \sum_{i\in\mathcal{I}} x_i = 0 .
    \end{align}
\end{subequations}
This can be seen as the ``economic equivalent'' to~\eqref{eq: dual epigraph}. That the dual optimal value $v^{\ast\ast}$ equals the optimal value of the convexified market follows again from \citet[Theorem~2.3]{lemarechal2001geometric}.

To obtain the economic version of~\Cref{lemma: extremepoints}, we will make use of the economic concept of demand sets.
The demand set of an agent under prices $\lambda$ is given by
\begin{equation}\label{eq: demand set}
    \mathcal{D}_i(\lambda) \; = \; \argmax_{x_i} \; u_i(x_i) - \lambda x_i 
\end{equation}
in the original market and by
\begin{equation*}
    \mathcal{D}_i^C(\lambda) \; = \; \argmax_{x_i} \; \cav{u_i}(x_i) - \lambda x_i 
\end{equation*}
in the convexified market. Both are related as follows.

\begin{lemma} \label{lemma: convex hull demand set}
For every $\lambda \in \mathbb{R}^m$ it holds that $\mathcal{D}^C_i(\lambda) = \co{\mathcal{D}_i(\lambda)}$.
\end{lemma}
\begin{proof}
See \citet[Theorem~3.4]{falk1969lagrange}.
\end{proof}

We can characterize a zero duality gap and thus the existence of an equilibrium through demand sets.

\begin{lemma}\label{lemma: equilibrium demand sets}
    A Walrasian equilibrium exists if and only if $0\in\sum_{i\in\mathcal{I}} \mathcal{D}_i(\lambda)$ at some price $\lambda$.
\end{lemma}
\begin{proof}
    Follows directly from the definition of demand sets and Walrasian equilibrium.
\end{proof}

Now, we can obtain the economic version of \Cref{lemma: extremepoints}. It shows that an equilibrium exists in the convexified market, and only a few agents are not allocated a bundle in their original demand sets~$\mathcal{D}_i$. We obtain it directly from the Shapley--Folkman lemma~\citep{starr1969quasi}, whose underlying argument we already used in \Cref{lemma: extremepoints}.

\begin{lemma}[Shapley--Folkman]\label{lemma: shapley-folkmann}
Let $A_i\subseteq\mathbb{R}^{m}$ for $i\in\mathcal{I}$. 
If 
$$x\in \co{\sum_{i\in\mathcal{I}} A_i} = \sum_{i\in\mathcal{I}} \co{A_i}$$ 
then there exists a subset $\mathcal{J}\subseteq\mathcal{I}$ with $\card{\mathcal{J}}\le m$ such that
$$x\in \sum_{i\in\mathcal{I}\setminus\mathcal{J}} A_i + \sum_{i\in\mathcal{J}} \co{A_i} ,$$
where the sum denotes the Minkowski addition of sets.
\end{lemma}

\begin{lemma}\label{lemma: shapley folkman demand sets}
In the convexified market, there exists an equilibrium $(x^\ast,\lambda^\ast)$ and a subset $\mathcal{J}\subseteq\mathcal{I}$ with $\card{\mathcal{J}}\le m$ such that $x_i^\ast \in \mathcal{D}_i(\lambda^\ast)$ for all $i\in\mathcal{I}\setminus\mathcal{J}$ and $x_i^\ast \in \co{\mathcal{D}_i(\lambda^\ast)}$ for all $i\in\mathcal{J}$.
\end{lemma}
\begin{proof}
By \Cref{assumption: market}(a), the duality gap of~\eqref{eq: convex welfare maximization} is zero and~\eqref{eq: dual welfare maximization} has a solution $\lambda^\ast$, so an equilibrium exists in the convexified market. By \Cref{lemma: equilibrium demand sets} applied to the convexified market and \Cref{lemma: convex hull demand set},
\begin{equation*}
    0 \in \; \sum_{i \in \mathcal{I}} \; \mathcal{D}^C_i(\lambda^\ast) \; = \; \sum_{i \in \mathcal{I}} \; \co{\mathcal{D}_i(\lambda^\ast)} ,
\end{equation*}
where $\sum$ is the Minkowski addition of sets. Applying \Cref{lemma: shapley-folkmann} to the demand sets $A_i=\mathcal{D}_i(\lambda^\ast)$ yields a set $\mathcal{J}\subseteq\mathcal{I}$ with $\card{\mathcal{J}}\le m$ and an allocation $x^\ast$ with $\sum_{i\in\mathcal{I}} x_i^\ast = 0$, $x_i^\ast\in\mathcal{D}_i(\lambda^\ast)$ for all $i\in\mathcal{I}\setminus\mathcal{J}$, and $x_i^\ast\in\co{\mathcal{D}_i(\lambda^\ast)}$ for all $i\in\mathcal{J}$. Since $x_i^\ast\in\mathcal{D}^C_i(\lambda^\ast)$ for all $i\in\mathcal{I}$, the tuple $(x^\ast,\lambda^\ast)$ is an equilibrium of the convexified market.
\end{proof}

In words, the demand sets $\mathcal{D}_i(\lambda^\ast)$ may be nonconvex, but at the convexified-market price $\lambda^\ast$, the market can be cleared by allocating to all but at most $m$ agents a bundle they actually demand. Only these at most $m$ agents, no more than there are commodities, must accept a convex combination of demanded bundles, and this number does not grow with the number of agents.

\subsection{Random Market}\label{subsec: probabilistic market}

So far, the analysis in this section has not used the randomness in \Cref{assumption: market}. We now use parts~(b) and~(c) and study the resulting \textit{random market}, populated by a mix of convex and nonconvex agents whose concave envelopes $\cav{u_i}$ are i.i.d.

By a similar argument as in the proof of \Cref{theorem: zero duality gap}, we can obtain a lower bound on the probability that a Walrasian equilibrium exists in this random market.
This is the economic equivalent of \Cref{theorem: zero duality gap}.

\begin{proposition}\label{proposition: equilibrium existence probability}
The probability that a Walrasian equilibrium exists is bounded below by:
\begin{equation*}
    \Pr(\textup{Equilibrium exists}) \ \ge \ \frac{\binom{\card{\mathcal{C}}}{m}}{\binom{\card{\mathcal{I}}}{m}} ,
\end{equation*}
where $\binom{n}{k} = \frac{n!}{k!(n-k)!}$ denotes the binomial coefficient.
\end{proposition}
\begin{proof}
Let $\mathcal{J}$ be a set of cardinality $m$ with the property in \Cref{lemma: shapley folkman demand sets}, chosen uniformly at random among all such sets if there are several. Since the concave envelopes $\cav{u_i}$ are i.i.d.\ by \Cref{assumption: market}(c), the symmetry argument of \Cref{lemma: uniform distribution} shows that $\mathcal{J}$ is uniformly distributed over all subsets of $\mathcal{I}$ with cardinality $m$.\footnote{To keep the economic argument transparent, this proof applies the symmetry argument of \Cref{lemma: uniform distribution} informally: whether $x_i^\ast\in\mathcal{D}_i(\lambda^\ast)$ holds depends on $u_i$ itself and not only on $\cav{u_i}$. A rigorous version selects $\lambda^\ast$ by a rule that treats all agents alike, e.g., as the solution to~\eqref{eq: dual welfare maximization} with the smallest Euclidean norm, and applies \Cref{lemma: shapley-folkmann} to the extreme points of the compact convex sets $\co{\mathcal{D}_i(\lambda^\ast)}$. These extreme points lie in $\mathcal{D}_i(\lambda^\ast)$ and depend only on $\cav{u_i}$, so that the argument of \Cref{lemma: extremepoints,lemma: uniform distribution} carries over. The same applies to the proof of \Cref{prop: market - non-zero duality gap}.} As in the proof of \Cref{theorem: zero duality gap}, the number of nonconvex indices $i\notin\mathcal{C}$ in $\mathcal{J}$ therefore follows a hypergeometric distribution.
That is, in an equilibrium in the convexified market, the number of nonconvex agents allocated a bundle outside their original demand set $\mathcal{D}_i(\lambda^\ast)$ follows a hypergeometric distribution.
But if no nonconvex agent is allocated a bundle outside their demand set, then an equilibrium must also exist in the original market (cf. \Cref{lemma: equilibrium demand sets}). Evaluating the hypergeometric distribution at 0 gives the result.
\end{proof}

For obtaining the economic equivalent of \Cref{theorem: non-zero duality gap}, we assume Lipschitz continuity of the value function of the welfare maximization problem~\eqref{eq: welfare maximization}:
\begin{equation}\label{eq: market value function}
    v(b) \; := \; \max_x \; \Big\{ \sum_{i\in\mathcal I} u_i(x_i) \; \Big| \; \sum_{i\in\mathcal I} x_i = b \Big\} .
\end{equation}
That is, $v(b)$ is the maximum welfare the market can attain if it must absorb an aggregate external imbalance $b$ rather than clear exactly. Note $v(0)=v^\ast$. We assume Lipschitz continuity of $v$ over the box
\begin{equation*}
    \mathcal{B} \; := \; \big\{ b\in\mathbb{R}^{m} \; \big| \; \norm{b}_\infty \le m \cdot 2\beta \big\} ,
\end{equation*}
where $\beta$ was the assumed deterministic bound on the compact domain $\dom{u_i}\subseteq[-\beta,\beta]^{m}$. 

What Lipschitz continuity of the value function means economically is the following. Let $\bar{x}$ be an optimal solution to the welfare maximization problem~\eqref{eq: welfare maximization}. If every agent produces/consumes $\bar{x}_i$ and we ``inject'' any additional aggregate supply or demand $b\in\mathcal{B}$ into the economy, welfare would not react disproportionately: there is no discontinuous jump for an infinitesimally small external supply or demand, and any perturbation $b\in\mathcal{B}$ changes total welfare by at most $L\norm{b}_\infty$ relative to $v^\ast$.

\begin{proposition}\label{prop: market - non-zero duality gap}
Assume that the value function of~\eqref{eq: welfare maximization} is Lipschitz continuous over $\mathcal{B}$ with constant $L$. Then
\begin{equation*}
    \Pr\big[v^{\ast\ast}-v^\ast \; \le \; (2(\gamma+L\beta)) \cdot t\big] \ \ge \ \sum_{r=0}^{t} \frac{\binom{\card{\mathcal{N}}}{r}\binom{\card{\mathcal{C}}}{m-r}}{\binom{\card{\mathcal{I}}}{m}}
\end{equation*}
for $t=0,1,\ldots,m$. In particular, the same bound holds for the lost opportunity cost $\Gamma(x',\lambda')$ of convex hull pricing, i.e., of any tuple $(x',\lambda')$ in which $x'$ solves~\eqref{eq: welfare maximization} and $\lambda'$ solves~\eqref{eq: dual welfare maximization}.
\end{proposition}
\begin{proof}
By \Cref{lemma: shapley folkman demand sets}, there is an equilibrium $(x^\ast,\lambda^\ast)$ in the convexified market and a set $\mathcal J\subseteq\mathcal I$ with $\card{\mathcal J}\le m$ such that $x_i^\ast\in\mathcal D_i(\lambda^\ast)$ for $i\notin\mathcal J$ and $x_i^\ast\in\co{\mathcal D_i(\lambda^\ast)}$ for $i\in\mathcal J$. Let $r:=\card{\mathcal J\cap\mathcal N}$: economically, $r$ is the number of nonconvex agents whom the convexified-market equilibrium asks to trade a bundle they cannot actually produce or consume.

Let us construct an alternative allocation $y$ that every agent can actually carry out, i.e., with $y_i\in\mathcal D_i(\lambda^\ast)$ for all $i$. For $i\notin\mathcal J\cap\mathcal N$, set $y_i:=x_i^\ast\in\mathcal D_i(\lambda^\ast)$. For $i\in\mathcal J\cap\mathcal N$, pick any $y_i\in\mathcal D_i(\lambda^\ast)$.

Because we replaced only the $r$ infeasible agents, the market no longer clears exactly; let $\delta:=\sum_{i\in\mathcal I} y_i$ denote the resulting aggregate imbalance. Using $\sum_{i\in\mathcal I} x_i^\ast=0$,
\begin{equation*}
    \delta \; = \; \sum_{i\in\mathcal I} y_i \; = \; \sum_{i\in\mathcal I} y_i - \sum_{i\in\mathcal I} x_i^\ast \; = \; \sum_{i\in\mathcal J\cap\mathcal N} (y_i-x_i^\ast) ,
\end{equation*}
since $y_i=x_i^\ast$ for every $i\notin\mathcal J\cap\mathcal N$. By definition of $\beta$ and $x_i^\ast\in\co{\mathcal D_i(\lambda^\ast)}$, each term satisfies $\norm{y_i-x_i^\ast}_\infty\le2\beta$, so $\norm\delta_\infty\le r2\beta\le m 2\beta$, i.e., $\delta\in\mathcal B$: re-routing the $r$ misallocated agents to feasible trades leaves only a bounded residual imbalance for the market to absorb. In particular, $(y_i)_{i\in\mathcal I}$ is a feasible allocation for~\eqref{eq: market value function} at balance $\delta$, so $v(\delta)\ge\sum_{i\in\mathcal I} u_i(y_i)$.

We now compare the welfare $\sum_{i\in\mathcal I} u_i(y_i)$ of $y$ to the convexified-market optimum $$v^{\ast\ast}=\sum_{i\in\mathcal I}\cav{u_i}(x_i^\ast).$$
From $\im{u_i}\subseteq[-\gamma,\gamma]$ and $\im{\cav{u_i}}\subseteq[-\gamma,\gamma]$ we get $u_i(y_i)\ge-\gamma\ge\cav{u_i}(x_i^\ast)-2\gamma$. 
Hence, since $\cav{u_i}(x_i^\ast) = u_i(x_i^\ast)$ for all $i\notin\mathcal J\cap\mathcal N$, it follows that
\begin{equation*}
    v(\delta) \; \ge \; \sum_{i\in\mathcal I} u_i(y_i) \; \ge \; \sum_{i\in\mathcal I} \cav{u_i}(x_i^\ast) - 2\gamma r \; = \; v^{\ast\ast} - 2\gamma r .
\end{equation*}
By Lipschitz continuity of $v$ over $\mathcal B\ni\delta$ with constant $L$ and $\norm\delta_\infty\le r2\beta$, it follows that
\begin{equation*}
    v^\ast \; = \; v(0) \; \ge \; v(\delta) - L\norm\delta_\infty \; \ge \; v^{\ast\ast} - 2\gamma r - Lr2\beta \; = \; v^{\ast\ast} - r(2(\gamma+L\beta)) .
\end{equation*}
Hence,
$$ v^{\ast\ast}-v^\ast\le r(2(\gamma+L\beta)).$$
Since the $\cav{u_i}$ are i.i.d.\ and $\mathcal J$ is uniform over size-$m$ subsets (as in the proof of \Cref{proposition: equilibrium existence probability}), $r=\card{\mathcal J\cap\mathcal N}$ follows $\mathrm{Hypergeom}(\card{\mathcal I},\card{\mathcal N},m)$. Since $r\le t$ implies $v^{\ast\ast}-v^\ast\le 2(\gamma+L\beta)t$, summing the hypergeometric probabilities from $r=0$ to $t$ gives the claimed bound. The statement on $\Gamma(x',\lambda')$ follows from \Cref{proposition: locs and duality gap}.
\end{proof}

In words, \Cref{prop: market - non-zero duality gap} tells us that it gets more and more likely that convex hull pricing yields a good approximate Walrasian equilibrium $(x',\lambda')$ as the number of convex agents in the market increases relative to the number of nonconvex agents. The quality of the approximation is measured by the lost opportunity cost $\Gamma(x',\lambda')$.

As in \Cref{subsec: value function}, we can ask when the assumed Lipschitz continuity of $v(b)$ over $\mathcal{B}$ actually holds. The economic equivalent of the condition in \Cref{lemma: lipschitz value function} is that for every additional external demand or supply $b\in\mathcal{B}$, and all bundles $(x_i)_{i\in\mathcal{N}}\in\prod_{i\in\mathcal{N}}\dom{u_i}$ that the nonconvex agents might demand or supply, there exist bundles $(x_i)_{i\in\mathcal{C}}\in\prod_{i\in\mathcal{C}}\relinterior{\dom{u_i}}$ so that
\begin{equation*}
    \sum_{i\in\mathcal{N}} x_i + \sum_{i\in\mathcal{C}} x_i = b .
\end{equation*}
That is, the convex agents can absorb any bundle that the nonconvex agents might demand, and can still rebalance the market after any admissible external shock $b\in\mathcal{B}$. Since $\dom{u_i}\subseteq[-\beta,\beta]^m$ and $\mathcal{B}=[-2m\beta,2m\beta]^m$, a sufficient condition is
\begin{equation}\label{eq: rebalancing}
    \big[-(\card{\mathcal{N}}+2m)\beta,\,(\card{\mathcal{N}}+2m)\beta\big]^m \; \subseteq \; \sum_{i\in\mathcal{C}} \relinterior{\dom{u_i}} .
\end{equation}
In the electricity auction of \Cref{sec: electricity}, this means that in every hour, the convex bids on both the buy and the sell side can absorb any position of the nonconvex bids.
By an argument similar to that in \Cref{subsec: value function}, it becomes increasingly likely that convex agents can rebalance the market as more of them enter.

\section{Electricity Auction}\label{sec: electricity}

A particular real-world market in which a Walrasian equilibrium is sought is the European day-ahead electricity auction. In a given country on a given day, agents $i\in\mathcal{I}$ submit a utility function $u_i$ by placing bids, and the nominated electricity market operator (NEMO) computes a tuple $(x',\lambda')$ that constitutes a Walrasian equilibrium whenever one exists. Given access to the utility functions $u_i$ and the tuple $(x',\lambda')$, we can characterize the duality gap of its welfare maximization problem~\eqref{eq: welfare maximization}.

In what follows, we study the empirical distribution of this duality gap across nine countries over 281 days in 2023, drawing on the theoretical results of \Cref{sec: duality gap,sec: equilibrium} to better understand our findings.

We describe the data in \Cref{subsec: data}, detail the bid formats that determine $u_i$ in \Cref{subsec: bids}, and introduce the resulting welfare maximization problem in \Cref{subsec: welfare maximization day-ahead}. 
Finally, in \Cref{subsec: duality gaps,subsec: discussion}, we report empirical results on duality gaps and equilibrium existence, and discuss them in light of the theory developed in \Cref{sec: duality gap,sec: equilibrium}.

\subsection{Data}\label{subsec: data}

We have access to commercial auction data from EPEX SPOT SE, which serves as NEMO for multiple European countries. Our data contains:
\begin{itemize}
    \item \textit{Nine countries:} Austria (AT), Belgium (BE), Finland (FI), France (FR), Germany (DE), Great Britain (GB), the Netherlands (NL), Poland (PL), and Switzerland (CH). 
    \item \textit{281 days in 2023:} The data set spans March 24 to December 31, 2023, excluding the time-shift days March 26 and October 29. 
\end{itemize}

The countries are interconnected through transmission lines, so their auctions are coupled. For all countries except Great Britain and Switzerland, this coupling is implicit and takes place within the Single Day-Ahead Coupling (SDAC). Great Britain and Switzerland are explicitly coupled: their day-ahead auctions clear at 11:00, before the SDAC clears at 12:00, so their cross-border exchanges with neighboring bidding zones are determined explicitly rather than jointly within the SDAC~\citep{euphemia2025}.

Data on earlier days in 2023 is also available, but does not include EUPHEMIA's bid acceptance rates, without which a meaningful equilibrium analysis is not possible; we therefore restrict attention to March 24 onward. We exclude the time-shift days, March 26 and October 29, because on those days the day-ahead auction featured 23 and 25 commodities instead of the regular 24, as the traded commodities are contracts for electricity delivery in each hour of the next day.

For each country and day, the auction data contains all submitted bids, their acceptance rates after clearing, and the resulting electricity prices. This is sufficient to compute $u_i$, $i\in\mathcal{I}$, and $(x',\lambda')$, and hence $\Gamma(x',\lambda')$. From \Cref{proposition: locs and duality gap}, we thus obtain an upper bound on the duality gap.

\subsection{Bid Formats}\label{subsec: bids}

On a given day in a given country, agents can trade contracts for constant power delivery in an hour $h\in\mathcal{H}=\{1,\ldots,24\}$ of the next day. That is, there are $m=24$ commodities, and the allocation $x_i$ to an agent $i$ lies in $\mathbb{R}^{24}$, with negative quantities denoting selling and positive quantities denoting buying.
Agents submit a utility function $u_i:\mathbb{R}^{24}\to\mathbb{R}\cup\{-\infty\}$ by placing bids. In the nine countries of our data set, agents use two types of bids: convex simple bids and nonconvex block bids~\citep{madani2015computationally,cacm_report_2023,euphemia2025,hubner2026package}.

\textit{Convex simple bids} consist of a tuple $(P_{1i},P_{2i},Q_{i},h)$, where $i$ indexes the bid and $h$ the hour for which the bid is submitted. $P_{1i},P_{2i}\in\mathbb{R}$ are two price points and $Q_i\in\mathbb{R}$ is a quantity. If the bid is a sell bid ($Q_i<0$), then $P_{1i}\le P_{2i}$ must hold, indicating increasing marginal production cost; if the bid is a buy bid ($Q_i>0$), then $P_{1i}\ge P_{2i}$ must hold, indicating decreasing marginal utility.
The bid acceptance rate is $\delta_{i}\in[0,1]$, and the allocated quantity is $x_i=Q_i\delta_i\mathbf{1}_h$ with $\mathbf{1}_h$ being the unit vector with a 1 at index $h$ and zero otherwise. The communicated utility function is
$$u_i(Q_i\delta_i\mathbf{1}_h) = P_{1i} Q_{i} \delta_i +  \frac{1}{2}(P_{2i} - P_{1i})Q_{i} \delta_i^2 ,$$
and is concave quadratic over the domain $\dom{u_i}=[0,Q_i\mathbf{1}_h]$. We denote by $\mathcal{C}_h\subseteq\mathcal{I}$ the set of all convex simple bids submitted for hour $h$.
The set of all convex bids is then $\mathcal{C} = \bigcup_{h\in\mathcal{H}}\mathcal{C}_h$.

\textit{Nonconvex block bids} consist of a tuple $(P_{i},Q_{i},M_i)$, where $i$ indexes the bid. $P_{i}\in\mathbb{R}$ is the price, $Q_i=(Q_{ih})_{h=1}^{24}\in\mathbb{R}^{24}$ is the quantity vector, and $M_i\in[0.01,1]$ is the minimum acceptance ratio (MAR).
The bid acceptance rate is $\delta_{i}\in\{0\}\cup[M_i,1]$, and the allocated quantity is $x_i=Q_i\delta_i$. The communicated utility function is
$$u_i(Q_i\delta_i) = P_i \sum_{h\in\mathcal{H}} Q_{ih} \, \delta_i$$
over the nonconvex domain
$$\dom{u_i} = \left\{ Q_i \delta_i \mid \delta_i \in \{0\}\cup[M_i,1] \right\}.$$
We denote by $\mathcal{N}\subseteq\mathcal{I}$ the set of all block bids. The acceptance rates $\delta_i$ of multiple block bids can be connected by logical relationships:
\begin{itemize}
    \item \textit{Ungrouped}: Each bid can be accepted or rejected independently of the others; the $\delta_i$ of ungrouped bids are not linked through any inequality or equality.
    \item As part of an \textit{exclusive group}: At most one bid in the group $\mathcal{E}\subseteq\mathcal{N}$ can be accepted, i.e., $\sum_{i\in\mathcal{E}} \delta_i \le 1$.
    \item \textit{Linked}: Bids are connected through ``if'' relationships or an ``if-and-only-if'' relationship~(so-called \textit{loop} bids). For instance, a relationship $\delta_i \le \delta_j$ indicates that $i$ can only be accepted if block bid $j$ is accepted.
\end{itemize}
We collect all inequalities and equalities that govern the logical relationships of the variables $(\delta_i)_{i\in\mathcal{N}}$ in the set $\mathcal{R}$, and write $(\delta_i)_{i\in\mathcal{N}}\in\mathcal{R}$ to indicate that they are enforced. Since these relationships couple the acceptance rates of different block bids, the welfare maximization problem is not separable across individual block bids. To relate it to the separable framework of \Cref{sec: duality gap,sec: equilibrium}, each exclusive group and each family of linked bids can be viewed as a single nonconvex agent.

EUPHEMIA supports further products that do not occur in these nine countries, such as complex orders with a minimum income condition and, in 2023, PUN orders in Italy~\citep{euphemia2025}. Their acceptance depends on prices, so they cannot be expressed through a utility function $u_i(x_i)$. Markets with such products are therefore not characterized by a welfare maximization problem, and our analysis does not apply to them.

\subsection{Optimization Problem}\label{subsec: welfare maximization day-ahead}

Given the two bid types defining the utility functions $u_i$, the welfare maximization problem~\eqref{eq: welfare maximization} of a given country on a given day can be written as:
\begin{subequations} \label{eq: winner determination}
\begin{align}
\max_{\delta_i} & \quad \sum_{h\in\mathcal{H}} \sum_{i\in\mathcal{C}_h} \Big( P_{1i} Q_{i} \cdot \delta_i +  \frac{1}{2}(P_{2i} - P_{1i})Q_{i} \cdot \delta_i^2 \Big) \; + \; \sum_{i\in\mathcal{N}} P_i \sum_{h\in\mathcal{H}} Q_{ih} \cdot \delta_i \label{eq: winner determination 0} \\
\text{s.t.} & \quad \sum_{i\in\mathcal{C}_h} Q_i \cdot \delta_i +  \sum_{i\in\mathcal{N}} Q_{ih} \cdot \delta_i = F_h \quad \forall h\in\mathcal{H} \label{eq: winner determination 1} \\
& \quad (\delta_i)_{i\in\mathcal{N}}\in\mathcal{R} \label{eq: winner determination 2} \\
& \quad \delta_i \in [0,1] \quad \forall i \in \mathcal{C}_h,  h\in\mathcal{H}  \label{eq: winner determination 3} \\
& \quad \delta_{i}\in\{0\}\cup[M_i,1] \quad \forall i \in \mathcal{N} \label{eq: winner determination 4}  .
\end{align}
\end{subequations}
The vector $F=(F_h)_{h\in\mathcal{H}}\in\mathbb{R}^{24}$ denotes the net import/export of this bidding zone in hours $h\in\mathcal{H}$. When the joint optimization problem over all coupled bidding zones is considered, $F$ is a decision variable; for a single bidding zone in isolation, it is instead a parameter, fixed at the value computed by EUPHEMIA for the joint problem.

\Cref{tab: bid formats} summarizes the use of bid formats by country, which in turn determines the parameters of the welfare maximization problem. The problem is dominated by its convex component: thousands of continuous variables $\delta_i, i\in\mathcal{C}_h$, account for the majority of the bid volume in the balance constraint~\eqref{eq: winner determination 1} (cf. \Cref{subtab: simple bid curves}), whereas far fewer nonconvex variables $\delta_i, i\in\mathcal{N}$, contribute considerably less volume (cf. \Cref{subtab: block bids}).

\begin{table}[t]
\centering
\small
\caption{Usage of bid formats.}\label{tab: bid formats}
    \begin{subtable}{\textwidth}
        \caption{Convex simple bids per day. Median of all 281 days.}
        \label{subtab: simple bid curves}
        \centering
        \begin{tabular}{c|ccccccccc}
            \hline
            & AT & BE & FI & FR & DE & GB & NL & PL & CH \\
            \hline 
            Simple bids [in thousands] & 5 & 9 & 7 & 13 & 19 & 23 & 18 & 25 & 32 \\
            Volume [GWh] & 217 & 158 & 310 & 780 & 1,599 & 387 & 318 & 225 & 367 \\
            Share of total bid volume [\%] & 93.3 & 66.2 & 89.8 & 81.6 & 80.6 & 63.6 & 68.2 & 99.9 & 96.8 \\
            \hline        
        \end{tabular}
    \end{subtable}
    \par\vspace{1em}
    \begin{subtable}{\textwidth}
        \caption{Nonconvex block bids per day. Median of all 281 days.}
        \label{subtab: block bids}
        \centering
        \begin{tabular}{c|ccccccccc}
            \hline
            & AT & BE & FI & FR & DE & GB & NL & PL & CH \\
            \hline
            Volume [GWh] & 15 & 79 & 37 & 183 & 385 & 222 & 152 & 0.4 & 12 \\
            Share of total bid volume [\%] & 6.7 & 33.8 & 10.2 & 18.4 & 19.4 & 36.4 & 31.8 & 0.1 & 3.2 \\
            Ungrouped bids & 43 & 73 & 133 & 79 & 225 & 398 & 102 & 1 & 38 \\
            Linked bids & 0 & 29 & 11 & 0 & 34 & 5 & 2 & 0 & 0 \\
            Loop bids & 0 & 0 & 0 & 0 & 0 & 38 & 0 & 0 & 0 \\
            Exclusive groups & 0 & 9 & 15 & 26 & 47 & 36 & 18 & 0 & 0 \\
            Median bids per group & -- & 24 & 11 & 24 & 18 & 24 & 24 & -- & -- \\
            \hline        
        \end{tabular}
    \end{subtable}
    \par\vspace{1em}
    \begin{subtable}{\textwidth}
        \caption{Minimum acceptance ratios (MAR) for all block bids over all 281 days in \%.}
        \label{tab: minimum acceptance ratio}
        \centering
            \begin{tabular}{cccccccccc}
                \hline
                 MAR & AT & BE & FI & FR & DE & GB & NL & PL & CH \\
                \hline
                0.01 & 0.0 & 2.1 & 0.7 & 3.9 & 16.1 & 69.1 & 10.0 & 0.0 & 0.1 \\
                (0.01, 0.5] & 1.4 & 6.1 & 11.5 & 0.4 & 0.1 & 6.5 & 2.7 & 0.4 & 0.0 \\
                (0.5, 1) & 0.0 & 8.0 & 10.2 & 20.7 & 2.5 & 8.4 & 16.1 & 0.2 & 0.0 \\
                1 & 98.6 & 83.8 & 77.6 & 75.1 & 81.3 & 16.0 & 71.3 & 99.4 & 99.9 \\
                \hline        
            \end{tabular}
        \end{subtable}
\end{table}

\subsection{Duality Gaps}\label{subsec: duality gaps}

The data on submitted bids, acceptance rates, and electricity prices allows us to compute $(x',\lambda')$ and $\Gamma(x',\lambda')$ for every country-day. Two properties of the mechanism producing $(x',\lambda')$ make this pair informative about the duality gap of~\eqref{eq: winner determination}. 
First, the algorithm EUPHEMIA, which computes $(x',\lambda')$, returns a Walrasian equilibrium whenever one exists; the duality gap is therefore zero exactly on the days on which $(x',\lambda')$ is an equilibrium~(\Cref{proposition: equilibrium existence duality gap}). 
Second, EUPHEMIA always returns an $x'$ satisfying $\sum_{i\in\mathcal{I}} x_i'=0$, so \Cref{proposition: locs and duality gap} applies, and the lost opportunity cost $\Gamma(x',\lambda')$ provides an upper bound on the duality gap on the remaining days. Note that $\Gamma(x',\lambda')$ need not equal the duality gap exactly: equality holds when $x'$ solves~\eqref{eq: welfare maximization} and $\lambda'$ solves~\eqref{eq: dual welfare maximization}, as is the case for the convex hull pricing mechanism~\citep{stevens2024some}, but this is not guaranteed for EUPHEMIA~\citep{madani2015computationally}.

We analyze each bidding zone in isolation, with its net position $F$ fixed at the value computed by EUPHEMIA. A zero lost opportunity cost in a bidding zone thus means that $(x',\lambda')$ is an equilibrium of that zone given its net position. In the coupled market, the lost opportunity cost is the sum of the zonal ones plus that of the transmission network, viewed as an agent that maximizes congestion rent over feasible flows. A zero lost opportunity cost in every zone is therefore necessary, but not sufficient, for an equilibrium of the coupled market. Likewise, at jointly optimal net positions, the duality gap of the coupled problem is at least the sum of the zonal duality gaps.

\begin{table}[t]
\centering
\small
\caption{Equilibrium existence and lost opportunity costs.}\label{tab:overview duality gaps}
    \begin{subtable}{\textwidth}
    \caption{Number of days where an equilibrium existed.}
    \label{tab:equilibria}
    \centering
    \begin{tabular}{l|ccccccccc}
        \hline
        & AT & BE & FI & FR & DE & GB & NL & PL & CH \\
        \hline
        Days without equilibrium & 42 & 173 & 115 & 235 & 244 & 266 & 221 & 2 & 50 \\
        Days with equilibrium & 239 & 108 & 166 & 46 & 37 & 15 & 60 & 279 & 231 \\
        \hline 
    \end{tabular}
    \end{subtable}
    \par\vspace{1em}
    \begin{subtable}{\textwidth}
    \centering
    \caption{Lost opportunity costs in thousand €.}
    \label{tab: lost opportunity cost absolute}
        \begin{tabular}{c|ccccccccc}
            \hline
             Quantiles & AT & BE & FI & FR & DE & GB & NL & PL & CH \\
            \hline
            25\% & 0.00 & 0.00 & 0.00 & 0.18 & 0.09 & 0.02 & 0.02 & 0.00 & 0.00 \\
            50\% & 0.00 & 0.14 & 0.00 & 1.28 & 0.84 & 0.15 & 0.96 & 0.00 & 0.00 \\
            75\% & 0.00 & 1.53 & 0.04 & 5.06 & 3.96 & 0.81 & 5.36 & 0.00 & 0.00  \\
            90\% & 0.01 & 7.37 & 0.67 & 17.12 & 11.49 & 2.31 & 17.41 & 0.00 & 0.03 \\
            95\% & 0.07 & 13.62 & 2.10 & 58.10 & 19.51 & 3.38 & 33.57 & 0.00 & 0.12 \\
            98\% & 1.94 & 23.81 & 13.62 & 80.98 & 32.38 & 5.30 & 55.21 & 0.00 & 0.34 \\
            100\% & 45.64 & 129.80 & 215.26 & 294.47 & 106.71 & 9.39 & 137.11 & 0.28 & 2.71 \\
            \hline        
        \end{tabular}
    \end{subtable}
    \par\vspace{1em}
    \begin{subtable}{\textwidth}
    \centering
    \caption{The auction's median welfare in million € over 281 days.}
    \label{tab: welfare}
        \begin{tabular}{cccccccccc}
            \hline
             & AT & BE & FI & FR & DE & GB & NL & PL & CH \\
            \hline
            Convex simple bids & 171 & 141 & 559 & 650 & 2,461 & 62 & 354 & 6 & 13 \\
            Nonconvex block bids & 0.18 & 0.69 & 0.12 & 6.26 & 13.74 & 1.28 & 0.37 & 0.00 & 0.01 \\
            \hline        
        \end{tabular}
    \end{subtable}
    \par\vspace{1em}
    \begin{subtable}{\textwidth}
    \centering
    \caption{Relative lost opportunity costs to total welfare in parts per million $(10^{-6})$.}
    \label{tab: lost opportunity cost relative}
        \begin{tabular}{c|ccccccccc}
            \hline
             Quantiles & AT & BE & FI & FR & DE & GB & NL & PL & CH \\
            \hline
            25\% & 0 & 0 & 0 & 0 & 0 & 0 & 0 & 0 & 0 \\
            50\% & 0 & 1 & 0 & 2 & 0 & 2 & 3 & 0 & 0 \\
            75\% & 0 & 10 & 0 & 9 & 2 & 13 & 18 & 0 & 0 \\
            90\% & 0 & 48 & 1 & 25 & 5 & 36 & 56 & 0 & 2 \\
            95\% & 0 & 87 & 4 & 92 & 8 & 55 & 94 & 0 & 8 \\
            98\% & 10 & 202 & 25 & 151 & 13 & 87 & 142 & 0 & 22 \\
            100\% & 292 & 1,342 & 385 & 406 & 55 & 177 & 423 & 51 & 448 \\
            \hline        
        \end{tabular}
    \end{subtable}
\end{table}

\Cref{tab:equilibria} reports, for each of the nine countries, the number of days on which an equilibrium existed given the country's net position, i.e., on which its welfare maximization problem had a zero duality gap. In countries such as Austria and Switzerland, the duality gap was zero on more than 80\% of days, whereas in Germany it was zero on only about 13\% of days. Since EUPHEMIA returns an equilibrium of the coupled market whenever one exists, and a zero lost opportunity cost in every zone is necessary for such an equilibrium, the SDAC as a whole had an equilibrium on at most 2 of the 281 days: those on which the lost opportunity cost was zero simultaneously in all seven SDAC countries of our data set.

\Cref{tab: lost opportunity cost absolute} reports the distribution of $\Gamma(x',\lambda')$, an upper bound on the absolute duality gap $v^{\ast\ast}-v^\ast$ (cf. \Cref{proposition: locs and duality gap}). \Cref{tab: welfare} reports the median objective value $\sum_{i\in\mathcal{I}} u_i(x_i')$, the welfare realized by allocation $x'$; note that $v^\ast \ge \sum_{i\in\mathcal{I}} u_i(x_i')$, since $x'$ need not be optimal for the welfare maximization problem. Comparing the two tables shows that the duality gap accounts for only a small fraction of the objective value.

\Cref{tab: lost opportunity cost relative} reports the distribution of $$\Gamma(x',\lambda') \; \big/ \;  \sum_{i\in\mathcal{I}} u_i(x_i'),$$ where $\sum_{i\in\mathcal{I}} u_i(x_i')$ is the objective value of the welfare maximization problem. Since $v^\ast \ge \sum_{i\in\mathcal{I}} u_i(x_i')$, this ratio bounds the relative duality gap $(v^{\ast\ast}-v^\ast)/v^\ast$ from above:
$$\frac{\Gamma(x',\lambda')}{\sum_{i\in\mathcal{I}} u_i(x_i')} \;\ge\; \frac{v^{\ast\ast}-v^\ast}{\sum_{i\in\mathcal{I}} u_i(x_i')} \;\ge\; \frac{v^{\ast\ast}-v^\ast}{v^\ast} .$$
The bounds are small on most days, indicating that a good approximate equilibrium was found frequently.

\subsection{Discussion}\label{subsec: discussion}

The submitted bids do not satisfy the i.i.d.\ assumption. They are not identically distributed, since a wind farm differs fundamentally from a thermal generator or a manufacturing plant, and they are not independent either, as bids on a given day are correlated through common factors such as weather, demand, and fuel prices. In particular, block bids are typically much larger than simple bids. This matters because our results are stated in terms of the numbers of convex and nonconvex functions, which under the i.i.d.\ assumption coincide in expectation with their volumes. In the data, the two differ substantially: evaluated with the numbers of bids in \Cref{tab: bid formats}, the bound of \Cref{proposition: equilibrium existence probability} exceeds 0.65 in France and Germany, whereas equilibria existed there on only 16\% and 13\% of the days. Extending the theory to volumes would require dropping the assumption of identically distributed functions, on which \Cref{lemma: uniform distribution} relies. We therefore compare the empirical results in \Cref{tab: bid formats,tab:overview duality gaps} and \Cref{fig: zero gap vs convexity} against \Cref{proposition: equilibrium existence probability,prop: market - non-zero duality gap} only \textit{qualitatively}, and measure the size of the convex part of the problem by its share of the bid volume.

\Cref{proposition: equilibrium existence probability} tells us that the likelihood of a zero duality gap increases with $\card{\mathcal{C}}$ relative to $\card{\mathcal{N}}$. \Cref{tab: bid formats} shows exactly this imbalance in every country: convex simple bids number in the thousands, against a few hundred nonconvex bids (each exclusive group and each family of linked bids counts as one nonconvex agent), and they account for the majority of the bid volume. Qualitatively, this is the setting in which \Cref{proposition: equilibrium existence probability} predicts that equilibrium existence, i.e., a zero duality gap, is likely, which is consistent with the frequent equilibria recorded in \Cref{tab:equilibria} for several countries. Across countries, the frequency of a zero duality gap increases with the convex share of the bid volume (\Cref{fig: zero gap vs convexity}), in line with the direction predicted by \Cref{proposition: equilibrium existence probability}.

The bound of \Cref{proposition: equilibrium existence probability} also shows that this imbalance alone need not make equilibria likely: the bound decreases quickly in the number of commodities, so that with $m=24$ even a moderate nonconvex share can make equilibria rare. A low frequency of equilibria is therefore not surprising in France and Germany, where block bids account for about 19\% of the bid volume (\Cref{tab: bid formats} and \Cref{fig: zero gap vs convexity}).

\Cref{prop: market - non-zero duality gap} predicts that for a problem with large $\card{\mathcal{C}}$ relative to $\card{\mathcal{N}}$, the duality gap distribution should concentrate near the lower end of its support. This matches the qualitative pattern in \Cref{tab: lost opportunity cost absolute,tab: lost opportunity cost relative}, where the relative lost opportunity cost is at most a few hundred parts per million on the large majority of days across most countries. Consistent with this, the bulk of the distribution sits closer to zero in countries with a higher convex share of the bid volume, i.e., in those on the right of \Cref{fig: zero gap vs convexity}.

Size is only one of the features of block bids that the i.i.d.\ assumption abstracts from. Block bids also span several hours, have minimum acceptance ratios that determine how far they are from convex (\Cref{tab: minimum acceptance ratio}), and are coupled through exclusive groups and links. \Cref{proposition: equilibrium existence probability,prop: market - non-zero duality gap} treat each block bid, or each exclusive group or family of linked bids, as one nonconvex function and thus ignore this structure. Exploiting it, for instance by measuring the nonconvexity of each block bid rather than bounding it uniformly, could yield sharper bounds.

\section{Conclusion}\label{sec: conclusions}

This article sought to explain a phenomenon observed in the European electricity auction: the mixed-integer welfare maximization problem frequently has a zero duality gap, so the auction is often in equilibrium.

Classical worst-case bounds on the duality gap cannot explain this, since they ignore the convex part of the problem. Our theory of duality gaps in partially nonconvex optimization, by contrast, shows that the phenomenon is to some extent expected: when the functions are random, zero duality gaps become increasingly likely the more convex functions outnumber nonconvex ones, and in the auction, convex bids far outnumber nonconvex ones.

Because the bids are neither independent nor identically distributed, this explanation is qualitative: it identifies a mathematical reason for the phenomenon rather than predicting its frequency in a given country.

More broadly, the phenomenon shows that zero duality gaps can be typical even when the worst-case bound is far from zero: when convex functions dominate, the worst case is atypical.

%============================================
% -----------  Bibliography ---------------
\bibliography{refs}
%============================================

%%%%%%%%%%%%%%%%%
\end{document}